\documentclass[11pt]{article}

\usepackage[final]{acl}

\usepackage{times}
\usepackage{latexsym}

\usepackage[T1]{fontenc}
\usepackage{amsmath}
\usepackage[utf8]{inputenc}

\usepackage{microtype}

\usepackage{inconsolata}

\usepackage{graphicx}
\usepackage{subcaption}
\usepackage{booktabs}
\usepackage{colortbl}
\usepackage{multirow}

\usepackage{amsfonts}

\usepackage{amssymb}

\definecolor{baselinecolor}{RGB}{235,243,252}
\definecolor{stochasticcolor}{RGB}{253,243,224}
\definecolor{historycolor}{RGB}{233,246,237}
\definecolor{baselineDark}{RGB}{35,80,135}
\definecolor{stochasticDark}{RGB}{145,90,20}
\definecolor{historyDark}{RGB}{35,110,65}

\newcommand{\sisd}[2]{\shortstack{$#1$\\[-1pt]{\scriptsize$\pm #2$}}}

\title{The Butterfly Defects: Tracing Cooperation in LLM Agent Networks}

\author{
  Maria Sampani \quad Maria Lymperaiou \quad Angeliki Dimitriou \\
  \textbf{Giorgos Filandrianos} \quad \textbf{Chrysoula Zerva} \quad \textbf{Giorgos Stamou} \\[4pt]
  National Technical University of Athens
}

\begin{document}
\maketitle
\begin{abstract}
As large language models (LLMs) increasingly operate as interconnected agents, a fundamental question emerges: \textit{does individual cooperation translate into collective cooperation?} We investigate this question through the Iterated Prisoner's Dilemma (IPD), moving from dyadic interactions to homogeneous and heterogeneous networks of open-weight LLMs. Our findings reveal that cooperation is highly sensitive to model composition and seemingly minor linguistic changes. Models that cooperate with identical peers can shift toward defection in mixed populations. A simple uncertainty cue can increase or decrease cooperation, redistribute costs, and affect unexposed agents several hops away. Even reversing action-label order can dramatically alter behavior, with uncertainty framing sometimes amplifying this sensitivity. Together, these findings challenge the notion of cooperation as a stable model-level property, revealing how strategic behavior depends on interaction partners, network positions, and the language used to frame decisions.
\end{abstract}

\section{Introduction}

As LLMs move from isolated tasks to interacting agents, a central question arises: what determines whether cooperation emerges, persists, or breaks down? In multi-agent environments, strategic behavior depends not only on individual capabilities but also on how rational agents adapt to one another under conflicting incentives. Game-theoretic social dilemmas, particularly the Iterated Prisoner's Dilemma (IPD), provide a controlled framework for stressing cooperation and strategic adaptation over repeated interactions \citep{nicer-than-humans,akata2025playing}. However, dyadic settings capture only part of this challenge: in decentralized networks, local decisions become interdependent, potentially shaping collective outcomes.
\begin{figure}[t!]
  \centering
  \includegraphics[width=1\linewidth]{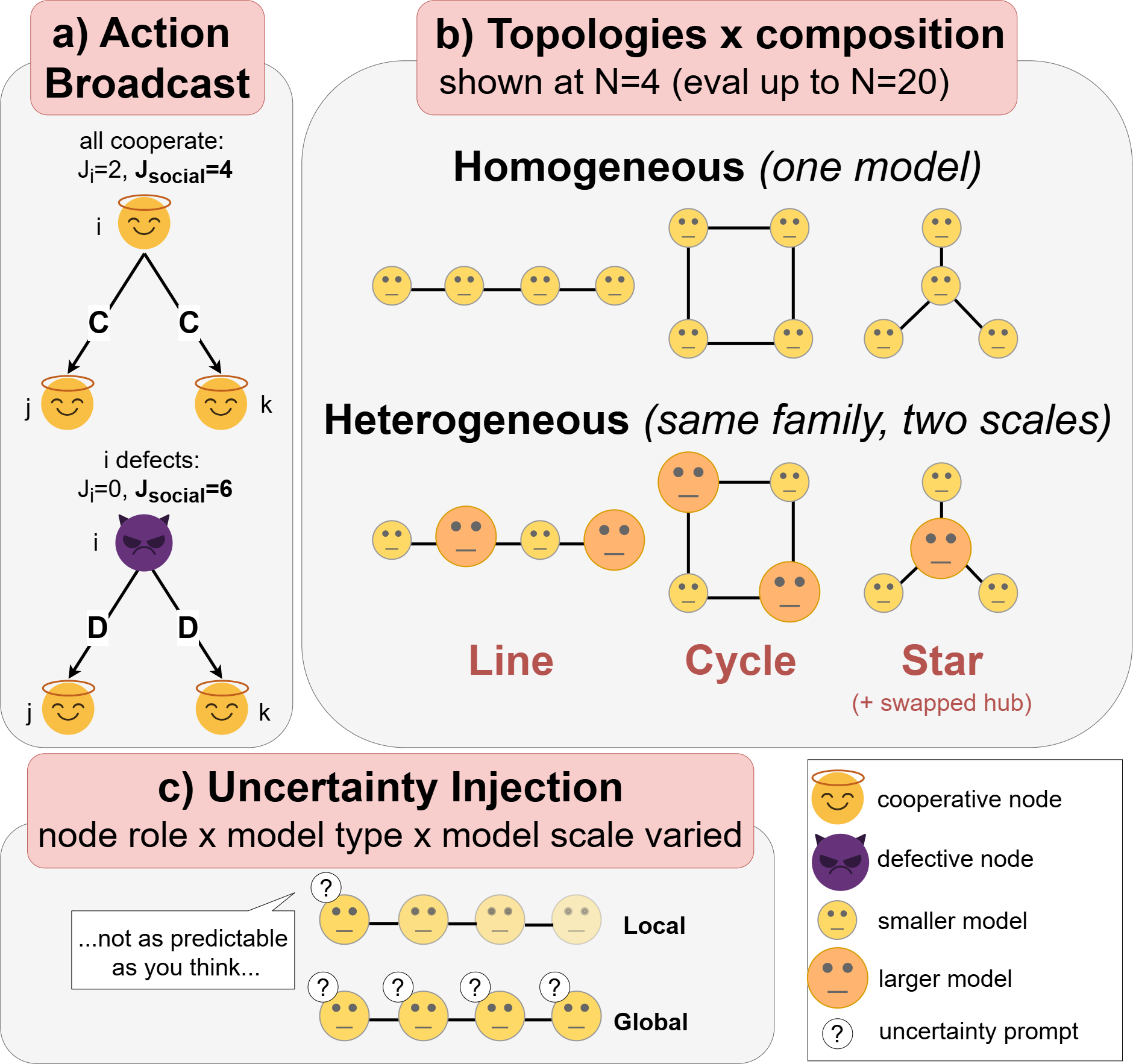}
  \caption{\textbf{Overview.} (a) Action broadcast to neighbors. (b) Topology types with homogeneous or heterogeneous model composition. (c) Uncertainty intervention types (illustrated on homogeneous graphs).}
  \label{fig:overview}
  
\end{figure}


One may intuitively expect cooperative agents to form a
cooperative network, yet this need not hold. Network topology
shapes who interacts with whom and the structural roles agents
occupy, while heterogeneous populations bring distinct strategic
tendencies into contact \citep{han2025static,qiu2025networkgames}.
These factors become particularly consequential in our setting,
where each agent selects one action shared across neighbors,
coupling simultaneous interactions. An agent that cooperates
against fixed opponents or identical peers may 
behave differently in a heterogeneous network. Understanding such shifts
requires considering graph structure, model composition, and
structural roles jointly.

Yet graph structure is only part of the story: LLM agents' strategies are also mediated by language. Contextual framing can shift behavior even when game incentives remain fixed \citep{loreheydari2024,korinek2026words}, while seemingly innocuous changes in response-option order can alter model choices \citep{zheng2024large,wei-etal-2024-unveiling}. In repeated interactions, such sensitivities may extend beyond the directly exposed agent, as its actions become part of its neighbors' action histories, potentially influencing subsequent decisions. Thus, a linguistic cue that does not relate to game structure may nevertheless reshape cooperation throughout the network, with consequences for both individual and collective welfare.

These observations raise a critical question: \textbf{when a single agent is influenced by a linguistic cue, how far can the consequences reach?} 

Prior work has examined networked cooperation and linguistic sensitivity, but how localized cues interact with model composition, structural roles, and action-label order---and affect unexposed agents and collective welfare---remains unclear.
To investigate this, we introduce a controlled networked IPD framework (Figure~\ref{fig:overview}), systematically varying topology, network size, model composition, and global or local exposure to uncertainty cues. We additionally examine whether strategic behaviors remain stable under semantically equivalent changes in action-label ordering in the prompt.


Our findings reveal three interconnected instability forms. Cooperation sustained among identical agents can deteriorate in heterogeneous networks; a cue delivered to a single agent can alter behavior several hops away or redistribute costs without changing unexposed agents' cooperation; and reversing action-label order can dramatically change cooperation, sometimes only after uncertainty framing is introduced. Together, these findings suggest that networked cooperation depends not merely on the agents involved, but also on whom they interact with and how the game is presented.

We first establish baseline behavior by pairing LLM agents with
fixed-policy opponents and placing them in homogeneous networks. Building
on these baselines, we investigate three novel research questions, evaluating
cooperation rates combined with resulting individual and social costs:

\begin{itemize}



    \item \textbf{RQ1: Heterogeneous networks.}
    When a smaller and a larger model from the same family share a network, does either shift from its homogeneous baseline? 

    \item \textbf{RQ2: Uncertainty cues.}
    (a) How does a cue questioning neighbor predictability affect cooperation and costs under \textit{global exposure}? (b) Under \textit{local exposure}, do behavioral changes extend to agents without the cue?

    \item \textbf{RQ3: Label order.}
    How much does swapping the order of C and D action labels in the instruction change behavior? Is the bias consistent in direction?
\end{itemize}

\section{Related work}
\paragraph{LLMs as Strategic Agents.}
LLMs have increasingly been studied as strategic agents in repeated social dilemmas. Prior work on IPD has examined cooperation, retaliation, and forgiveness against artificial and human opponents, revealing model-dependent strategic tendencies \citep{nicer-than-humans,akata2025playing}. Other studies investigate LLM-designed strategies and their evolutionary performance \citep{willis2025cooperation}, as well as cooperation under varying incentives, horizons, and framing conditions \citep{pal2026cooperation}. Together, these findings highlight discrepancies between game-theoretic incentives and LLM behavior.

\paragraph{Networked Cooperation among LLM Agents.}
Network topology influences the evolution of cooperation \citep{ohtsuki2006simple}, motivating its study in LLM populations. \citet{han2025static} found that static network structures do not reliably stabilize cooperation, while \citet{qiu2025networkgames} examined how connectivity and personality assignments influence cooperation, allowing agents to select actions separately for each neighbor. \citet{dearruda2026collective} identified discrepancies between aggregate cooperation and individual decision rules relative to human behavior, whereas \citet{decurto2026theory} investigated topology-dependent cooperation and externally imposed defection shocks. Related studies show that communication topology affects collective performance \citep{li-etal-2024-improving-multi} and information propagation \citep{shen-etal-2025-understanding}. Building on these findings, we examine model composition, structural roles, and localized interventions in repeated networks where agents broadcast a single action to all neighbors.

\paragraph{Linguistic Sensitivity and Behavioral Propagation.}
LLM strategic decisions are influenced not only by game incentives but also by contextual framing \citep{loreheydari2024,korinek2026words}. Sensitivity to option ordering and response labels \citep{zheng2024large,wei-etal-2024-unveiling} further raises concerns about behavioral stability under semantically equivalent prompts. \citet{ye2026stopdrawingscientificclaims} demonstrate that minor changes in simulation design can substantially alter cooperation, emphasizing the need for robustness audits. While linguistic sensitivity and information propagation have been investigated separately, we examine their interaction in networked IPD, assessing whether behavioral effects of uncertainty cues extend beyond exposed agents, affect individual and collective costs, and depend on action-label ordering.

\section{Method}

\subsection{Game-Theoretic Environment}
\label{sec:game_environment}

We study repeated pairwise Prisoner's Dilemma interactions
between agents connected through an undirected graph
$G=(V,E)$, where $V=\{1,\ldots,n\}$ denotes the set of
agents and $\mathcal{N}(i)$ the neighborhood of agent $i$.
At each round $t$, every agent $i$ selects a single action:
\begin{equation}
    a_i^t \in \{C,D\},
\end{equation}
where $C$ denotes cooperation and $D$ denotes defection. Crucially, $a_i^t$ is node-level, \emph{not} edge-specific:
it is broadcast to all neighbors $j\in\mathcal{N}(i)$ (Fig.~\ref{fig:overview}(a)).
 This constraint ensures that the network game does not decompose into independent pairwise games.

The pairwise stage game is defined by a cost function
$c(a_i,a_j)$, giving the years in prison assigned to agent $i$ when interacting with agent $j$ (lower is better), with the following cost matrix:
\begin{equation}
\begin{array}{c|cc}
a_i \backslash a_j & C & D \\
\hline
C & (1,1) & (3,0) \\
D & (0,3) & (2,2)
\end{array}
\end{equation}
where each entry reports the costs of agents $i$ and $j$,
respectively. 
For either action of the opponent,
\begin{equation}
    c(D,a_j) < c(C,a_j),
    \qquad a_j\in\{C,D\}
\end{equation}
so defection strictly dominates cooperation in the one-shot
game. Consequently, mutual defection is the unique Nash equilibrium of each pairwise game, and the all-defect
action profile is the unique Nash equilibrium of the induced network game.

Although the one-shot game incentives are fixed,
decisions are coupled over time through local interaction
histories. At round $t$, agent $i$ observes its own and
its neighbors' actions from at most the previous ten
rounds:
\begin{equation}
h_i^t =
\left(
a_i^\tau, a_{\mathcal{N}(i)}^\tau
\right)_{\tau=\max(1,t-10)}^{t-1}.
\label{eq:history}
\end{equation}
where $a_{\mathcal{N}(i)}^\tau =
(a_j^\tau)_{j\in\mathcal{N}(i)}$.
A possibly stochastic policy maps interaction histories
to probability distributions over actions:
\begin{equation}
    \pi_i : \mathcal{H}_i \rightarrow \Delta(\{C,D\}).
\end{equation}
Eqs.~\ref{eq:individual-score} and \ref{eq:network-score} define the individual cumulative cost of agent $i$ and the network-level social cost over $T$ rounds, respectively.
\begin{equation}
    J_i =
    \sum_{t=1}^{T}
    \sum_{j\in\mathcal{N}(i)}
    c\!\left(a_i^t,a_j^t\right).
    \label{eq:individual-score}
\end{equation}
\begin{equation}
    J_{\mathrm{social}} =
    \sum_{t=1}^{T}
    \sum_{\{i,j\}\in E}
    \left[
        c\!\left(a_i^t,a_j^t\right)
        +
        c\!\left(a_j^t,a_i^t\right)
    \right].
    \label{eq:network-score}
\end{equation}
Individual optimality, though, is distinct from social efficiency. Consider a single round of the three-node example in Fig.~\ref{fig:overview}(a): mutual cooperation gives agent $i$ a cost of $J_i=2$ and a social cost of $J_{\mathrm{social}}=4$. If $i$ instead broadcasts $D$ while both neighbors cooperate, its individual cost falls to $J_i=0$, but the social cost rises to $J_{\mathrm{social}}=6$. Reducing individual cost can thus increase collective cost. More generally, $(C,C)$ Pareto-dominates $(D,D)$ and minimizes the joint
cost on an edge, so the all-cooperate profile minimizes $J_{\mathrm{social}}$ despite not corresponding to a Nash equilibrium of the one-shot game. This tension between individual rationality and social efficiency defines the Prisoner's Dilemma \citep{banerjee2007reaching}, and it allows us to examine whether LLM agents tend toward individually defensive outcomes, socially efficient cooperation, or unstable intermediate behaviors as a function of interaction history, network
structure, model composition, uncertainty cues, and prompt formulation.

\subsection{Agent Design}
\label{sec:agent_design}

We consider two types of agents: \textit{adaptive} LLM agents, the focus of
this work, and \textit{fixed-policy} baselines, deterministic and stochastic
strategies used as controlled reference opponents in dyadic experiments
(App.~\ref{app:strategies}, Table~\ref{tab:strategies}).

Each LLM agent $i$ is instantiated from a pretrained language
model and prompted with the game rules, the outcome costs, and
its serialized local history $h_i^t$ at round $t$. It is asked to select the action leading to the best
long-term strategic outcome; the prompt \emph{does not} instruct it to minimize its own cumulative cost, maximize social welfare, cooperate, or defect, so strategic tendencies emerge from the model's interpretation of the game rather than from
the instructions.
The model is queried once per round and must return a single
action $a_i^t\in\{C,D\}$, 
broadcast to its neighbors. 
Outputs other than $C$ or $D$ are marked \textit{invalid} and excluded from evaluation. Full prompts are in App.~\ref{app:prompts}.


\subsection{Network Topologies}
\label{sec:network_topologies}

Since strategic outcomes depend on graph constraints \citep{vaknin2025solving} and cooperation is shaped by the underlying interaction graph \citep{ohtsuki2006simple}, we restrict our analysis to three canonical topologies (Fig.~\ref{fig:overview}(b)). 
These canonical graphs capture distinct connectivity patterns and offer interpretable structural roles. This allows systematic comparisons of behavior across topology, network size, and model assignments, without assuming that observed differences can be attributed to any single structural factor.
Specifically, evaluated topologies are: 
(i) \textbf{Line} ($n\geq 2$): a path with endpoints of degree 1 and internal nodes of degree 2. It isolates the asymmetry between boundary and internal positions; 
(ii) \textbf{Cycle} ($n>2$): a closed ring where every node has degree 2. 
All roles are structurally equivalent, providing a symmetric reference; 
(iii) \textbf{Star} ($n>3$): a hub of degree $n-1$ connected to $n-1$ leaves of degree 1. It isolates centrality and concentrates influence at the hub.

\subsection{Experimental Conditions}

\paragraph{Dyadic reference policies.}
We first evaluate each LLM agent against fixed-policy opponents
(App.~\ref{app:strategies}, Table~\ref{tab:strategies}) covering unconditional cooperation (Always Cooperate, AC) or defection (Always Defect, AD), stochastic play (Random, RND), and history-dependent adaptation (Tit-For-Tat, TFT). These interpretable reference behaviors expose each model's default tendencies before moving to LLM-only networks.


\paragraph{Network size and model composition.}
In networks, we vary the number of agents to test whether behavioral patterns remain stable, attenuate, or propagate as the network grows. We also vary model composition (Fig.~\ref{fig:overview}b): \textit{homogeneous} networks instantiate every node from the same model, while \textit{heterogeneous} networks pair a small and a large model from the same family, alternating along lines and cycles and placed at the hub or the leaves of stars, in both directions. Network sizes and model pairings are given in Sec.~\ref{sec:experimental_setup}.

\paragraph{Uncertainty cue.}
\label{sec:uncertainty}

We examine how agents respond to a textual cue questioning the predictability of their neighbors’ strategies. We append the following sentence to the prompt:

\begin{quote}
\small
\textit{``The other players' strategy is not as predictable as
you think, and it is defined from an unknown distribution.''}
\end{quote}

Payoffs, rules, topology, and generation settings remain unchanged, and no external randomization is introduced: only the agent's textual information about neighbor predictability is modified. Our intervention therefore tests behavioral sensitivity to uncertainty-related framing rather than directly measuring agents’ uncertainty or introducing additional stochasticity into the environment.



We compare a \textit{baseline} condition, in which the cue is omitted, with two exposure conditions (Fig.~\ref{fig:overview}c). Under global exposure, every agent receives the cue. Under local exposure, exactly one designated agent receives it, while the others receive the original prompt and are not informed of the intervention. 
Varying the exposed agent’s model and structural role allows us to examine whether behavioral changes remain confined to that agent or extend to unexposed agents.
All other prompt components
and experimental parameters are held fixed across the three conditions.

\subsection{Evaluation}
\label{sec:evaluation}


Our primary behavioral measure is the \textbf{Cooperation Rate} of agent
$i$, computed over rounds in which the agent produces a valid
action:
\begin{equation}
\mathrm{CoopRate}_i =
\frac{
    \sum_{t=1}^{T}
    \mathbb{1}[a_i^t=C]
}{
    \sum_{t=1}^{T}
    \mathbb{1}[a_i^t\in\{C,D\}]
}
\label{eq:coop_rate}
\end{equation}
For homogeneous networks, we additionally report network-level
cooperation by averaging $\mathrm{CoopRate}_i$ across 
nodes. For heterogeneous networks, we report cooperation separately for the small- and large-model node groups. When positional roles are relevant, i.e. hubs and leaves in star graphs, we also retain role-specific $CoopRate$.


Cooperation rate describes agents' action preferences,
while Average Cost (social and individual) measures the resulting strategic
outcomes. Since invalid actions do not define valid game
outcomes, we compute costs only over jointly evaluable
rounds. Let
\[
\mathcal{T}_{\mathrm{valid}}
=
\{t \in \{1,\ldots,T\} :
a_i^t \in \{C,D\},\ \forall i \in V\},
\]
with $\widetilde{T}=|\mathcal{T}_{\mathrm{valid}}|$.
We report the mean social cost per node and jointly
evaluable round:
\begin{equation}
\overline{J}_{\mathrm{social}}
=
\frac{1}{N\widetilde{T}}
\sum_{t\in\mathcal{T}_{\mathrm{valid}}}
\sum_{\{i,j\}\in E}
\left[
c(a_i^t,a_j^t)+c(a_j^t,a_i^t)
\right].
\end{equation}
Here, $N=|V|$, and individual costs use the same jointly
evaluable rounds. Lower costs indicate better welfare,
although they also reflect network connectivity.
Cooperation rates are undefined when an agent produces
no valid actions, and costs when $\widetilde{T}=0$;
these values are reported as unavailable. Estimates
based on fewer than $T$ rounds have reduced coverage.


Our default prompt lists cooperation before defection
($\mathrm{C}_{\mathrm{first}}$). Unless otherwise stated,
the results preceding Sec.~\ref{sec:label-order} refer
to this ordering.
To evaluate sensitivity to label-ordering,
we define the \textbf{Sensitivity Index} (SI).
We introduce a semantically equivalent prompt variant
that lists defection first ($\mathrm{D}_{\mathrm{first}}$),
holding all other prompt components and experimental
conditions fixed. 
For each configuration $\kappa$ and seed $s$, SI is the absolute difference in cooperation rate between the two variants (Eq. \ref{eq:sensitivity_index}), averaged over matched seeds $s\in\mathcal{S}$ (Eq. \ref{eq:mean_sensitivity}).
\begin{equation}
\hskip -0.135in \mathrm{SI}_{\kappa,s}
=
\left|
\mathrm{CoopRate}_{\kappa,s}^{
    \mathrm{C}_{\mathrm{first}}
}
-
\mathrm{CoopRate}_{\kappa,s}^{
    \mathrm{D}_{\mathrm{first}}
}
\right|
\label{eq:sensitivity_index}
\end{equation}
\begin{equation}
\overline{\mathrm{SI}}_{\kappa}
=
\frac{1}{|\mathcal{S}|}
\sum_{s\in\mathcal{S}}
\mathrm{SI}_{\kappa,s}
\label{eq:mean_sensitivity}
\end{equation}
Here, $\mathcal{S}$ denotes the set of matched experimental
seeds, controlling for stochastic variation. SI ranges from 0 to 1; higher
values indicate greater sensitivity in cooperation
rates. In network experiments, cooperation rates are
averaged over the evaluated node group before
computing SI. Thus, SI captures aggregate
cooperation shifts rather than individual policy
changes: opposing node-level responses may cancel,
and identical cooperation rates need not imply
identical action trajectories. A configuration
$\kappa$ denotes a model--opponent pair in dyadic
experiments or a combination of topology, network
size, model composition, uncertainty condition,
and evaluated node group in networks.

\section{Experimental Setup}
\label{sec:experimental_setup}
\paragraph{Models.}
We evaluate open-weight instruction-tuned LLMs from three families: Qwen2.5-Instruct, LLaMA-3.1-Instruct, and Mistral-Instruct (identifiers in App.~\ref{app:model-ids}).

\paragraph{Network Configurations.}
Line graphs use $n\in\{3,4,10,20\}$, while cycle and star graphs use
$n\in\{4,10,20\}$. Homogeneous networks are built for each of the eight models.
Heterogeneous networks use four within-family small--large pairs: Qwen2.5-3B
and Qwen2.5-7B, each with Qwen2.5-32B; Llama-3.1-8B with Llama-3.1-70B; and
Mistral-7B with Mixtral-8$\times$7B. Every configuration is run under both
$\mathrm{C}_{\mathrm{first}}$ and $\mathrm{D}_{\mathrm{first}}$ and under the
baseline, global, and local uncertainty conditions.

\paragraph{Runs.} 
Each configuration is repeated with three seeds. Since decoding is greedy
($\tau=0$), runs with deterministic opponents typically yield identical
trajectories, so repetitions mainly confirm reproducibility. Dyadic experiments
against stochastic opponents (\textsc{RND}, \textsc{URNDP}) use ten seeds to
cover variation in sampled actions. 
Each interaction lasts $T=100$ rounds, and
agents are exposed at most the ten most recent rounds in their prompt. The game horizon is
not disclosed, which avoids end-game effects from a known final round.

\section{Results and Analysis}

\subsection{Dyadic behavior}

\begin{figure}[t]
    \centering
    \includegraphics[width=\linewidth]
    {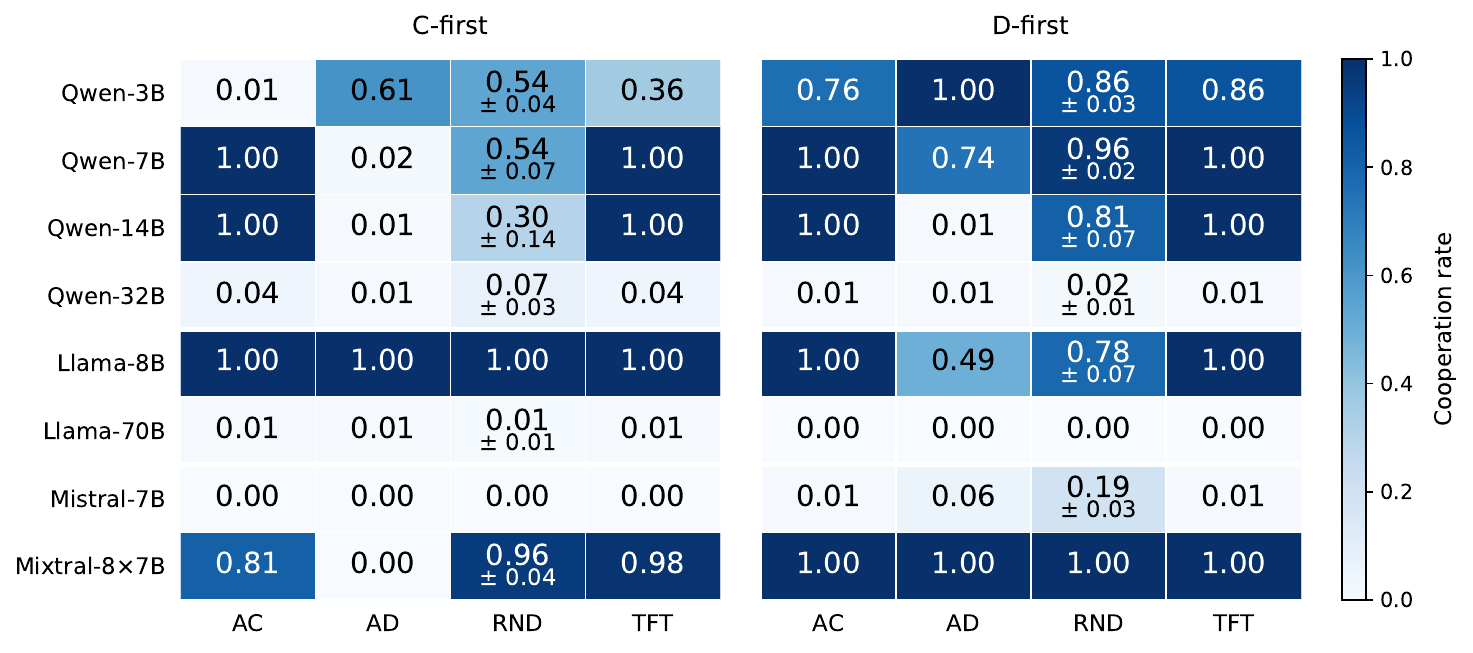}
    \caption{
    Dyadic cooperation against fixed opponents
    under $C_{\mathrm{first}}$ and $D_{\mathrm{first}}$
    prompts. Cells report means across seeds;
    non-zero sample standard deviations appear below.
    }
    \label{fig:dyadic-cooperation}
\end{figure}

Dyadic behavior \emph{varies} across models, ranging from
largely opponent-invariant dispositions to strong
opponent dependence, with no monotonic relationship
between model scale and cooperation.
Fig.~\ref{fig:dyadic-cooperation} reports cooperation rates
against 
AC, AD, RND, and TFT. For example, Qwen-14B's cooperation
is 1.00 against AC and TFT but only 0.01 against AD,
illustrating strong opponent dependence. In contrast,
Llama-70B predominantly defects against all four opponents, illustrating a largely opponent-invariant disposition. 
Scale is not monotonically associated with cooperation:
Qwen's cooperation
against AC rises from 0.01 at 3B to 1.00 at 7B and 14B,
then falls to 0.04 at 32B. Corresponding individual average costs are reported in Fig.~\ref{fig:dyadic-cost} in the appendix.
Results for the remaining four policies are provided in App.~\ref{app:dyadic-results}. 
In the following subsections, we examine whether these observed tendencies persist when opponents are themselves LLM agents and interactions are coupled through a shared action across neighbors.

\subsection{Homogeneous networks.}



Dyadic behavior need not carry over to interactions among LLM agents:
some models exhibit substantial behavioral differences between
dyadic fixed-policy interactions and homogeneous LLM networks.
Homogeneous networks reveal an additional layer of complexity: beyond model identity, cooperation \emph{also depends} on graph structure and network size.
Some models exhibit little variation across topologies
and network sizes: Qwen-3B consistently defects,
whereas Qwen-14B and Mixtral remain nearly fully
cooperative in configurations with available estimates.

Still others respond in a structure-specific way:
Increasing $N$
from 4 to 20 reduces Llama-70B's cooperation from 0.96
to 0.05 in cycles but increases it from 0.51 to 0.91
in stars
(App.~\ref{app:homogeneous-results},
Table~\ref{tab:homogeneous_cooperation}). Corresponding social costs are reported in Table~\ref{tab:homogeneous_social_cost}.
If cooperation already varies with structure and size when all agents are identical, a natural next question is whether it remains stable when models of different sizes interact.




\subsection{Heterogeneous networks.}
\textit{\textbf{RQ1}: Yes, heterogeneous interaction can shift the cooperation of both
the smaller and the larger model away from their homogeneous baselines, with
the magnitude and direction depending on the pairing and network configuration.}

\begin{figure*}[t]
    \centering
    \includegraphics[width=\textwidth]{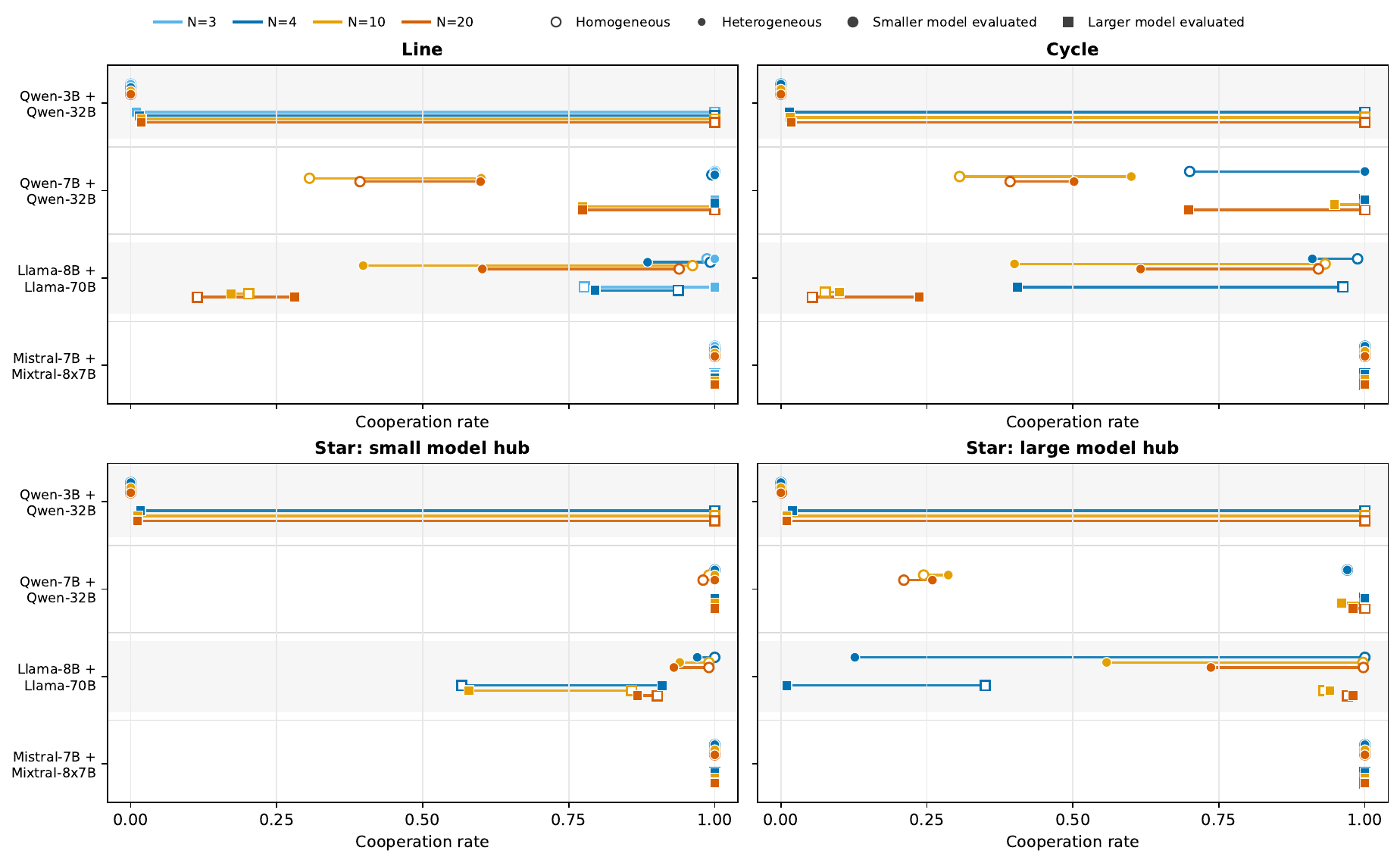}
    \caption{
    Cooperation shifts from homogeneous to heterogeneous graphs
    without the uncertainty cue ($C_{first}$). Lines connect matched settings;
    star baselines match hub/leaf roles. Values are averaged equally across seeds.
    }
    \label{fig:heterogeneous_shifts}
\end{figure*}
\begin{figure}[t]
    \centering
    \includegraphics[width=\columnwidth]{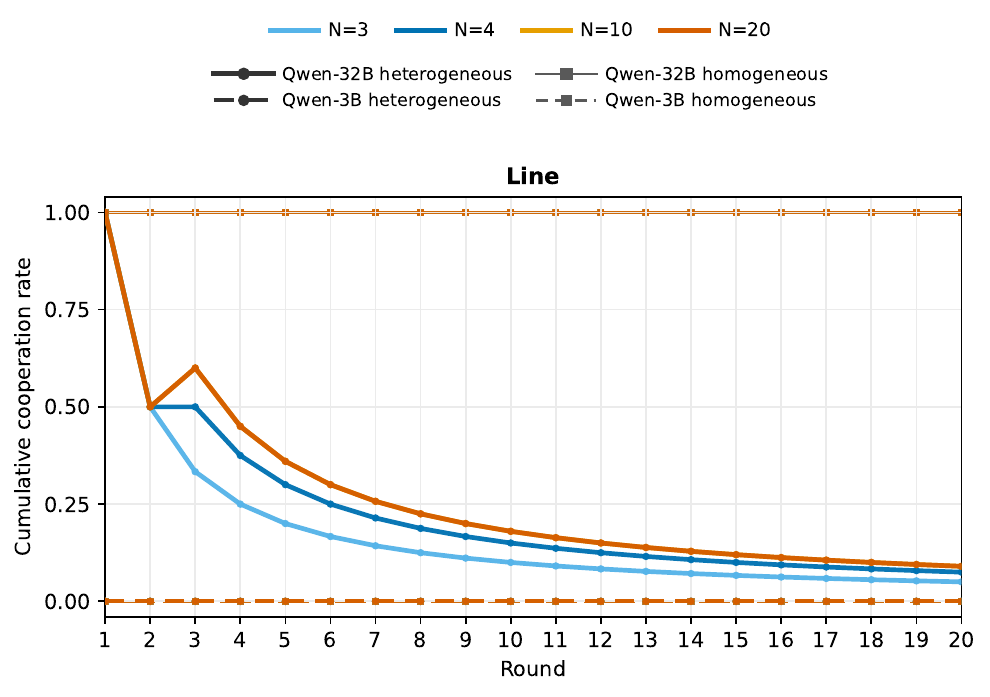}
    \caption{
    Cumulative cooperation 
    per round for selected Qwen configurations (cue-off).}
    \label{fig:qwen_line_trajectories}
\end{figure}

This shift is exemplified by pronounced asymmetry exhibited in the Qwen-3B/32B pairing.
Fig.~\ref{fig:heterogeneous_shifts} compares each model's $CoopRate$ with its homogeneous reference.
Qwen-3B remains predominantly defecting, while Qwen-32B
loses much of the cooperation it exhibited among identical
peers. 
The per-round cumulative $CoopRate$ trajectories in
Fig.~\ref{fig:qwen_line_trajectories} show how this
departure unfolds: Qwen-32B initially cooperates, as in its homogeneous
baseline, but its cumulative cooperation subsequently
falls toward Qwen-3B's near-zero level, diverging from
its homogeneous reference.
Thus, 
heterogeneous interaction 
can
disproportionately disrupt one model's cooperative behavior
while leaving the other's unchanged. Cycle and star trajectories in
App.~\ref{app:het-results-unc-off}.


Other pairings exhibit smaller or differently directed
shifts. In larger lines and cycles, pairing Qwen-32B with
Qwen-7B reduces its cooperation less strongly than pairing
it with Qwen-3B, while Qwen-7B becomes more cooperative
than in its homogeneous reference.

As for role-specific observations, in heterogeneous stars (Fig.~\ref{fig:heterogeneous_shifts} bottom row), behavior also depends on which
model occupies the hub. 
For example, the Llama-8B/70B pairing illustrates strong dependence on
hub assignment: an 8B hub remains nearly fully
cooperative with 70B leaves, whereas mean cooperation across the $N-1$ 8B leaves decreases with a 70B hub.


\subsection{Uncertainty cues.}
\textit{\textbf{RQ2(a):}
    Global exposure alters cooperation non-uniformly and redistributes costs:
social cost can fall while individual cooperators pay more.}

\begin{figure}[t]
    \centering
    \includegraphics[width=0.85\columnwidth]{
        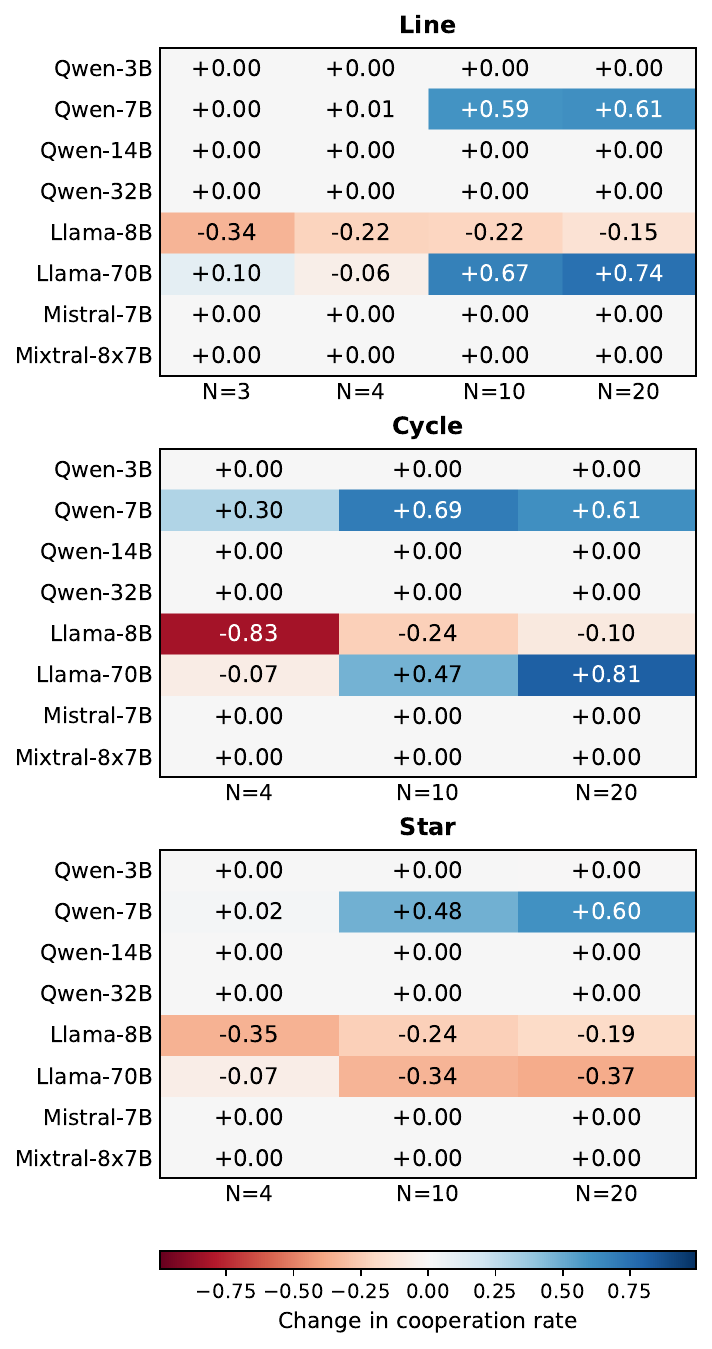
    }
    \caption{
    Global cue shifts on homogeneous cooperation under
    $\mathrm{C}_{\mathrm{first}}$:
    $\Delta CoopRate$ (cue on $-$ off),
computed per matched seed and averaged across seeds
    }
    \label{fig:homogeneous_uncertainty_effects}
\end{figure}

\paragraph{Global cue exposure in homogeneous networks.}
Fig.~\ref{fig:homogeneous_uncertainty_effects} shows that
global cue exposure does not \emph{uniformly} promote or suppress
cooperation.
Several models retain their baseline behavior, while others
respond in ways that depend on network size and topology.
Qwen-7B generally becomes more cooperative, particularly
in larger networks, whereas Llama-8B becomes less cooperative
across configurations. Llama-70B illustrates how network
structure can reverse the response: cooperation increases
in larger lines and cycles but decreases in stars.
Thus, the same cue can leave behavior unchanged, reinforce
cooperation, or weaken it. 
We hypothesize that models differ in reading uncertainty as a \emph{risk}
warranting defection or as a reason to \emph{build trust}.
Corresponding social-cost changes
are reported in App.~\ref{app:global-uncertainty-homogeneous},
Fig.~\ref{fig:homogeneous_social_uncertainty_effects}.

\paragraph{Global cue exposure in heterogeneous networks.}
Global uncertainty cues can change the collective welfare while redistributing individual costs (App.~\ref{app:het-results-unc-on}, Table~\ref{tab:heterogeneous-global-uncertainty-summary}). Social cost decreases in both Qwen pairings, but increases slightly in Llama-8B/70B, where models exhibit similar cooperation shifts. In Qwen-3B/32B networks, increased Qwen-32B cooperation benefits Qwen-3B and reduces social cost while increasing Qwen-32B's own cost. 
This redistribution follows from the dilemma itself: unreciprocated
cooperation lowers both partner's and social cost at the
cooperator's expense.

Network position also shapes who benefits: in a 20-node
Qwen-7B/32B star, increased leaf cooperation (0.259 to 0.842)
reduces the 32B hub's individual cost (46.76 $\to$ 25.00),
despite little change in the hub's cooperation.
Leaf costs, however, rise
(Table~\ref{tab:heterogeneous-global-uncertainty-configurations}(a)).
Since the hub interacts with every leaf, it accumulates this benefit
across its edges, whereas each leaf bears the cost of its single
interaction.
The cue can also pull cooperation back toward the homogeneous baseline
(Table~\ref{tab:heterogeneous-global-uncertainty-configurations}(b)).

\textit{\textbf{RQ2(b):}
When only one agent receives the cue, behavioral changes can extend to unexposed peers.}

\paragraph{Local cue exposure.}

\begin{table}[t]
\centering
\small
\setlength{\tabcolsep}{3pt}
\begin{tabular}{@{}lllrr@{}}
\toprule
Network & Topology & Exposed node & $d$ & $\Delta C(d)$ \\
\midrule
Llama-8B/70B & Line  & 70B endpoint & 10 & $+0.010$ \\
             & Line  & 8B interior  & 9  & $+0.015$ \\
             & Cycle & 70B node     & 10 & $-0.020$ \\
             & Cycle & 8B node      & 9  & $-0.015$ \\
\addlinespace
Llama-8B  & Line & Endpoint      & 5 & $-0.010$ \\
Llama-70B & Line & Endpoint      & 3 & $+0.010$ \\
Qwen-7B   & Line & Interior node & 3 & $+0.065$ \\
\bottomrule
\end{tabular}
\caption{
Farthest observed nonzero $\Delta CoopRate (\Delta C)$ in selected
$N=20$ networks under $\mathrm{C}_{\mathrm{first}}$ local
cue exposure. $d$ is shortest-path distance from the exposed
node; $\Delta C(d)$ averages changes from matched cue-off configurations.
}
\label{tab:localized_uncertainty_distance}
\end{table}



Single-agent exposure can change the cooperation of unexposed agents
(\emph{behavioral spillovers}) or their costs (\emph{outcome spillovers}), depending on the configuration
(App.~\ref{app:local-uncertaity},
Fig.~\ref{fig:localized_response_extent}).
The two need not co-occur: in Qwen-3B/32B cycles, exposing Qwen-32B increases its
cooperation by 0.4 without changing unexposed cooperation.
Yet unexposed individual costs fall by
0.265, the exposed agent's cost rises by 0.8, and social cost per node
falls by 0.107 (Table~\ref{tab:localized_uncertainty_roles}). 
As under global exposure, unreciprocated cooperation
benefits others at the cooperator's expense.

The exposed agent's role and model shape both magnitude and 
direction of the response.
In Llama-8B/70B stars, exposing the 70B hub
raises 
unexposed $CoopRate$ by 0.411 (vs 0.009 for an 8B leaf);
with roles reversed, exposing the 8B hub lowers it by 0.108 (vs
0.022 for a 70B leaf).
Hub exposure thus produces larger
shifts, 
in a direction set by the hub’s model.
(Table~\ref{tab:localized_uncertainty_roles}).

Localized effects can also reach distant agents.
At $N=20$, exposing one 70B agent in Llama-8B/70B networks changes cooperation ten hops
away (increase in lines, decrease in cycles, 
where ten hops is
the farthest node).
Homogeneous reach is shorter: five hops for Llama-8B and three for Llama-70B  (endpoint exposure), and three
for Qwen-7B (interior exposure)
(Table~\ref{tab:localized_uncertainty_distance}).
Notably, the same Llama-70B exposure travels farther in a mixed line than among identical peers.
Reach and direction of localized exposure depend on model composition, topology, and exposure location.

\subsection{Label order}
\label{sec:label-order}
\textit{\textbf{RQ3}: 
Label reversal can dramatically alter cooperation without a consistent positional preference.}

Choosing a single action-label order can leave its influence
on behavior unexamined. By supplementing our default
$\mathrm{C}_{\mathrm{first}}$ prompt with the semantically
equivalent $\mathrm{D}_{\mathrm{first}}$, we reveal
cooperation shifts that would remain \emph{hidden} under a single
formulation. 




Models differ in which position they favor. Under
$\mathrm{D}_{\mathrm{first}}$, Mixtral's cooperation against AD rises
from 0.00 to 1.00, favoring the second-listed action, whereas Llama-8B
favors the first-listed action, cooperating more under
$\mathrm{C}_{\mathrm{first}}$. Sensitivity also depends on the setting:
Mixtral is maximally sensitive against AD and STFT (SI $=1.00$) but
insensitive in homogeneous networks
(Tables~\ref{tab:dyadic-si-by-opponent} and~\ref{tab:homogeneous_cooperation}).

Label order also qualifies structural and composition effects. Qwen-7B's
decline in cooperation with network growth appears only under
$\mathrm{C}_{\mathrm{first}}$; under $\mathrm{D}_{\mathrm{first}}$ it
remains high. Conversely, Qwen-32B's cooperation in 20-node stars
collapses from 1.00 to 0.02 when switching to
$\mathrm{D}_{\mathrm{first}}$ (Table~\ref{tab:homogeneous_cooperation}).
The heterogeneity effect from RQ1 is likewise order-dependent: in larger
Qwen-7B/32B lines and cycles, Qwen-7B exceeds its homogeneous cooperation
only under $\mathrm{C}_{\mathrm{first}}$, as both settings are fully
cooperative under $\mathrm{D}_{\mathrm{first}}$
(Fig.~\ref{fig:heterogeneous_shifts_dfirst}; App.~\ref{app:het-results-unc-off}).

Finally, label order changes the cue's effects. Qwen-14B
is almost insensitive to order without the cue (SI $=0.001$) but becomes
almost fully sensitive under global exposure (SI $=0.990$)
(Table~\ref{tab:homogeneous-si-uncertainty-on}). In heterogeneous networks, the cue can either amplify
or attenuate order sensitivity: Qwen-32B's mean SI
increases with Qwen-3B, whereas Qwen-7B's decreases 
when paired with Qwen-32B
(Table~\ref{tab:heterogeneous-si-global-uncertainty}). Under local exposure,
the same intervention even reverses its consequences: exposing the
70B hub in Llama-8B/70B stars raises mean unexposed cooperation by 0.411
and lowers social cost by 0.458 under $\mathrm{C}_{\mathrm{first}}$, but
lowers cooperation by 0.144 and raises social cost by 0.228 under
$\mathrm{D}_{\mathrm{first}}$ (Table~\ref{tab:localized_uncertainty_roles}).
Label order can thus alter both behavior and welfare, so conclusions about LLM cooperation \emph{should} be checked
across prompt formulations.

\section{Conclusion}
We studied cooperation among LLM agents in the Iterated Prisoner's
Dilemma, from dyadic play to homogeneous and heterogeneous networks.
Cooperation proved fragile: it shifted with population composition and network position, a
single uncertainty cue caused shifts to distant unexposed agents and redistributed
costs, and even reversing the action-label order changed outcomes.
Cooperative agents do not guarantee a cooperative network, so multi-agent
LLM systems should be evaluated across the populations, structures, and
prompt formulations in which they will be deployed.

\section*{Limitations}
Our findings concern the evaluated models and a controlled
networked IPD protocol. We consider three canonical topologies,
networks of up to 20 agents, and selected within-family model
pairings. Each agent broadcasts one action to all neighbors,
without communication or neighbor-specific decisions.
Interactions last 100 rounds with a ten-round history window.
These choices enable controlled comparisons but limit
generalization to larger, more diverse, or differently
structured multi-agent environments.

We operationalize uncertainty-related framing through one
textual cue about neighbor predictability, without measuring
subjective uncertainty or changing environmental stochasticity. Label-order reversal tests one form of prompt
sensitivity rather than robustness to prompting more broadly.
Greedy decoding and repeated seeds primarily assess
reproducibility in deterministic settings. Finally, metrics
exclude invalid outputs and therefore characterize behavior
conditional on valid actions; cost estimates based on few
jointly valid rounds provide limited evidence about outcomes
over the full interaction.

\section*{Ethical Considerations}

Although this study involves no human participants or
real-world decision-making, its findings could inform
attempts to manipulate the behavior of deployed LLM
agent networks. Such applications warrant appropriate
safeguards and independent robustness evaluation.
Additionally, experiments involving large language
models incur computational and environmental costs,
which were not quantified in this study.


\bibliography{custom}

\appendix

\section{Background in Prisoners' Dilemma}
\label{app:game_theory}

\begin{figure*}[ht!]
    \centering
    \includegraphics[width=0.8\linewidth]{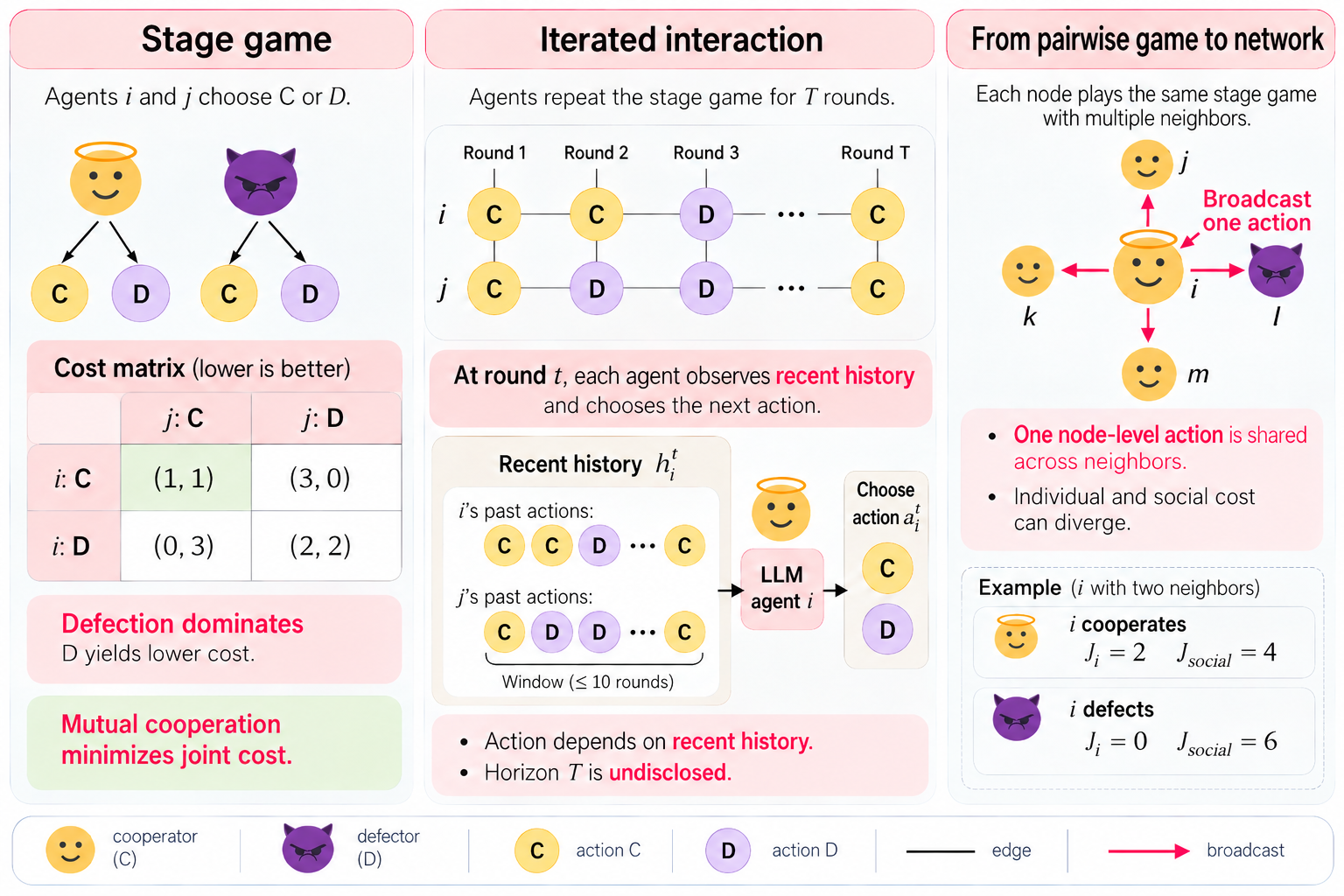}
    \caption{ A detailed overview of the 
        \textbf{Iterated Prisoner's Dilemma (IPD)} game setting.
        The pairwise stage game defines the costs of
        cooperation and defection. Agents repeatedly select
        actions based on recent interaction histories.
        In networked IPD, each agent broadcasts a single
        action to all neighbors, coupling individual
        decisions and collective outcomes.
    }
    \label{fig:ipd-background}
\end{figure*}

Figure~\ref{fig:ipd-background} illustrates the stage game,
its repeated interactions, and the extension to networks.

\paragraph{Cost convention.}
The four Prisoner's Dilemma outcomes are conventionally
termed \emph{temptation} (defect against cooperation),
\emph{reward} (mutual cooperation), \emph{punishment}
(mutual defection), and the \emph{sucker's outcome}
(cooperate against defection).
Denoting their costs by
$T_{\mathrm{PD}}$, $R_{\mathrm{PD}}$,
$P_{\mathrm{PD}}$, and $S_{\mathrm{PD}}$,
respectively, we have
\begin{equation}
\begin{aligned}
T_{\mathrm{PD}} = 0
&< R_{\mathrm{PD}} = 1 \\
&< P_{\mathrm{PD}} = 2 \\
&< S_{\mathrm{PD}} = 3.
\end{aligned}
\end{equation}
The ordering is reversed relative to the conventional
payoff formulation because lower costs are preferable.
The total cost per edge is $2R_{\mathrm{PD}}=2$ under
mutual cooperation, $T_{\mathrm{PD}}+S_{\mathrm{PD}}=3$
under asymmetric play, and $2P_{\mathrm{PD}}=4$
under mutual defection.

\paragraph{Network externalities.}
Let $d_i=|\mathcal{N}(i)|$ denote the degree of agent $i$.
If agent $i$ switches from $C$ to $D$ in round $t$,
holding all other agents' actions fixed, its immediate
cost decreases by $d_i$, while each neighbor's cost
increases by $2$. Thus, the changes in the round-$t$
contributions to individual and social cost are
\begin{equation}
\Delta J_i^{(t)}=-d_i,\qquad
\Delta J_{\mathrm{social}}^{(t)}=d_i.
\end{equation}
Defection therefore creates a negative externality:
it benefits the deviating agent at the immediate
expense of its neighbors and collective welfare.
The magnitude of this effect scales with node degree,
linking individual incentives to structural roles.

\paragraph{Repeated-game considerations.}
These immediate incentives need not determine behavior
over repeated interactions. An agent's current action
becomes part of its neighbors' observed histories,
potentially triggering reciprocity, retaliation, or
other strategic responses in subsequent rounds.
In a finitely repeated Prisoner's Dilemma with a
commonly known horizon and fully rational
cost-minimizing agents, backward induction predicts
defection in every round. This serves only as a
theoretical reference here: LLM agents are not told
the horizon or instructed to solve for an equilibrium.
Their strategies are instead examined empirically
through their responses to interaction histories.

\section{Fixed Strategies}
\label{app:strategies}
Table~\ref{tab:strategies} defines the fixed-policy opponents used in the dyadic experiments. Memoryless policies provide reference responses to
unconditional cooperation, defection, and stochastic behavior,
while history-dependent policies test interactions involving
reciprocity, retaliation, and outcome-dependent switching.
These opponents follow predefined rules throughout the game,
allowing the LLM's response to be examined against a controlled
interaction partner.
\begin{table*}[ht!]
\vskip -0.1in
\centering
\small
\begin{tabular}{l p{5.5cm} p{5.3cm}}
\toprule
\textbf{Strategy} & \textbf{Description} & \textbf{Policy Definition} \\
\hline

\rowcolor{baselinecolor}
\textbf{Always Cooperate (AC)} 
& Always plays cooperation ($C$). 
& $\pi_i(h_i^t) = C \;\forall t$ \\

\rowcolor{baselinecolor}
\textbf{Always Defect (AD)} 
& Always plays defection ($D$). 
& $\pi_i(h_i^t) = D \;\forall t$ \\

\rowcolor{stochasticcolor}
\textbf{Random (RND)} 
& Uniform random action selection. 
& $\pi_i(h_i^t) = C$ w.p. $0.5$, $D$ w.p. $0.5$ \\

\rowcolor{stochasticcolor}
\textbf{Uncertain Random (URNDP)} 
& Biased random strategy ($p=0.2$). 
& $\pi_i(h_i^t) = D$ w.p. $p$, $C$ w.p. $1-p$ \\

\rowcolor{historycolor}
\textbf{Tit-for-Tat (TFT)} 
& First $C$, then mirrors opponent. 
& $\pi_i(h_i^1)=C$; $\pi_i(h_i^t)=a_j^{t-1}$ \\

\rowcolor{historycolor}
\textbf{Suspicious Tit-for-Tat (STFT)} 
& First $D$, then mirrors opponent. 
& $\pi_i(h_i^1)=D$; $\pi_i(h_i^t)=a_j^{t-1}$ \\

\rowcolor{historycolor}
\textbf{Grim Trigger (GRIM)} 
& Permanent $D$ after first opponent defection. 
& $\pi_i(h_i^t)=C$ if no prior $D$; otherwise $D$ \\

\rowcolor{historycolor}
\textbf{Win-Stay, Lose-Shift (WSLS)}
& Repeats action if previous cost is favorable, otherwise switches.
& $\pi_i(h_i^t)=a_i^{t-1}$ if outcome $\in\{R_{\mathrm{PD}},T_{\mathrm{PD}}\}$; \newline else $\neg a_i^{t-1}$ \\

\bottomrule
\end{tabular}
\vskip -0.04in
\caption{
Strategies used as fixed opponent policies.
\textcolor{baselineDark}{Baseline strategies} include memoryless deterministic policies, 
\textcolor{stochasticDark}{Stochastic} introduce controlled randomness, and
\textcolor{historyDark}{History-dependent strategies} condition actions on prior interactions.
}
\label{tab:strategies}
\end{table*}
\vskip -0.15in

\section{Model Identifiers}
\label{app:model-ids}
Our experiments consider the following model families: Qwen 2.5 \cite{qwen2025qwen25technicalreport}, LLaMa 3.1  \cite{grattafiori2024llama3herdmodels} and Mistral \cite{jiang2024mixtralexperts}.

Table~\ref{tab:model-ids} lists the exact model identifiers used in the experiments.

\begin{table}[t]
\centering
\small
\begin{tabular}{ll}
\toprule
Family & Model Identifier \\
\midrule
Qwen2.5 & \texttt{Qwen/Qwen2.5-3B-Instruct} \\
Qwen2.5 & \texttt{Qwen/Qwen2.5-7B-Instruct} \\
Qwen2.5 & \texttt{Qwen/Qwen2.5-14B-Instruct} \\
Qwen2.5 & \texttt{Qwen/Qwen2.5-32B-Instruct} \\
\midrule
LLaMA-3.1 & \texttt{meta-llama/Llama-3.1-8B-Instruct} \\
LLaMA-3.1 & \texttt{meta-llama/Llama-3.1-70B-Instruct} \\
\midrule
Mistral & \texttt{mistralai/Mistral-7B-Instruct-v0.2} \\
Mistral & \texttt{mistralai/Mixtral-8x7B-Instruct-v0.1} \\
\bottomrule
\end{tabular}
\caption{Exact model identifiers used in the experiments.}
\label{tab:model-ids}
\end{table}

\section{Prompts}
\label{app:prompts}

For reproducibility, this appendix reports the prompt templates used in the experiments.

The dyadic prompt describes interaction with a single opponent,
whereas the network prompt specifies that one action is applied
simultaneously to all neighbors and that individual cost sums
across these interactions. Label-order comparisons reverse the
presentation of the available actions while preserving their
meanings and the game rules. Uncertainty interventions add the
cue to all agents or to a single exposed agent, according to the
experimental condition.

\paragraph{Comparability of experimental settings.}
Dyadic experiments involve fixed-policy opponents and a two-player prompt, whereas network experiments involve interacting LLM agents and a network-specific prompt. Consequently, differences between these settings cannot be attributed exclusively to network structure, as opponent type and prompt formulation also change. Dyadic results therefore serve as reference behaviors rather than controlled estimates of the isolated effect of network embedding.

\subsection{Prompt for dyadic fixed-policy opponents}
\label{app:prompt-twoplayer}

The following prompt template was used in two-player experiments against fixed baseline agents:

\begin{quote}
\small
\begin{verbatim}
You are playing the Iterated Prisoner's Dilemma.

Game rules:
- Both cooperate (C): 1 year each
- You defect (D), opponent cooperates (C): 
  0 years for you,
  3 years for opponent
- Both defect (D): 2 years each

Previous moves:
You: {history_self}
Opponent: {history_opp}

Instructions:
- Reply only with 'C' (cooperate) or 'D' (defect).
- Do not add explanations or extra text.
- Choose the move that leads to the best 
  strategic outcome across all rounds, 
  considering both players' previous moves.
- Do not follow moral or social advice; 
  base your decision solely on strategy.
- FINAL ANSWER:
\end{verbatim}
\end{quote}


\paragraph{Action-label ordering.}
We evaluate two semantically equivalent prompt variants,
$\mathrm{C}_{\mathrm{first}}$ and
$\mathrm{D}_{\mathrm{first}}$, which reverse the
presentation order of cooperation and defection
without changing their meanings or game incentives.
The response-option instructions are:

\begin{quote}
\small
\textbf{Dyadic prompt:}

$\mathrm{C}_{\mathrm{first}}$:
\texttt{Reply only with 'C' (cooperate) or 'D' (defect).}

$\mathrm{D}_{\mathrm{first}}$:
\texttt{Reply only with 'D' (defect) or 'C' (cooperate).}

\medskip
\textbf{Network prompt:}

$\mathrm{C}_{\mathrm{first}}$:
\texttt{Reply only with 'C' or 'D'.}

$\mathrm{D}_{\mathrm{first}}$:
\texttt{Reply only with 'D' or 'C'.}
\end{quote}

\subsection{Prompt for network topologies}
\label{app:prompt-network}

The following prompt template was used for agents in network topologies (line, cycle, and star):

\begin{quote}
\small
\begin{verbatim}
{graph_info_block}
You are playing the Iterated Prisoner's Dilemma 
against all of your neighbors simultaneously.

Rules:
- Each round, you choose C (cooperate) 
  or D (defect).
- You simultaneously play a Prisoner's Dilemma 
   with each neighbor.
- Your total cost = sum of the costs from 
  all neighbors.

Pairwise PD outcomes:
- (C,C) = 1 year each
- (D,C) = 0 years for you, 3 for neighbor
- (C,D) = 3 years for you, 0 for neighbor
- (D,D) = 2 years each

Your recent actions:
You: {history_self}

Neighbor histories (per neighbor):
{round_history}

{extra_hint}

Instructions:
- Reply only with 'C' or 'D'.
- Do not add explanations or extra text.
- Choose the move that leads to the best long-
  term 
strategic outcome
  across multiple neighbors in the network.
- FINAL ANSWER:
\end{verbatim}
\end{quote}

Here, \texttt{\{graph\_info\_block\}} specifies the node identity and neighborhood, \texttt{\{history\_self\}} denotes the agent's recent actions, \texttt{\{round\_history\}} contains the observed actions of neighboring nodes, and \texttt{\{extra\_hint\}} is used to inject uncertainty cues.

\paragraph{Handling invalid responses.}
When an LLM produces an action other than $C$ or $D$, the simulation continues without retrying or replacing the response. Invalid outputs are retained in the interaction history and may influence subsequent decisions. For evaluation, cooperation rates exclude invalid actions for the corresponding agent, while cost estimates include only rounds in which all agents produce valid actions (Sec.~\ref{sec:evaluation}).

\section{Dyadic Results}
\label{app:dyadic-results}
\begin{figure}[tbp]
    \centering
    \includegraphics[width=\linewidth]
    {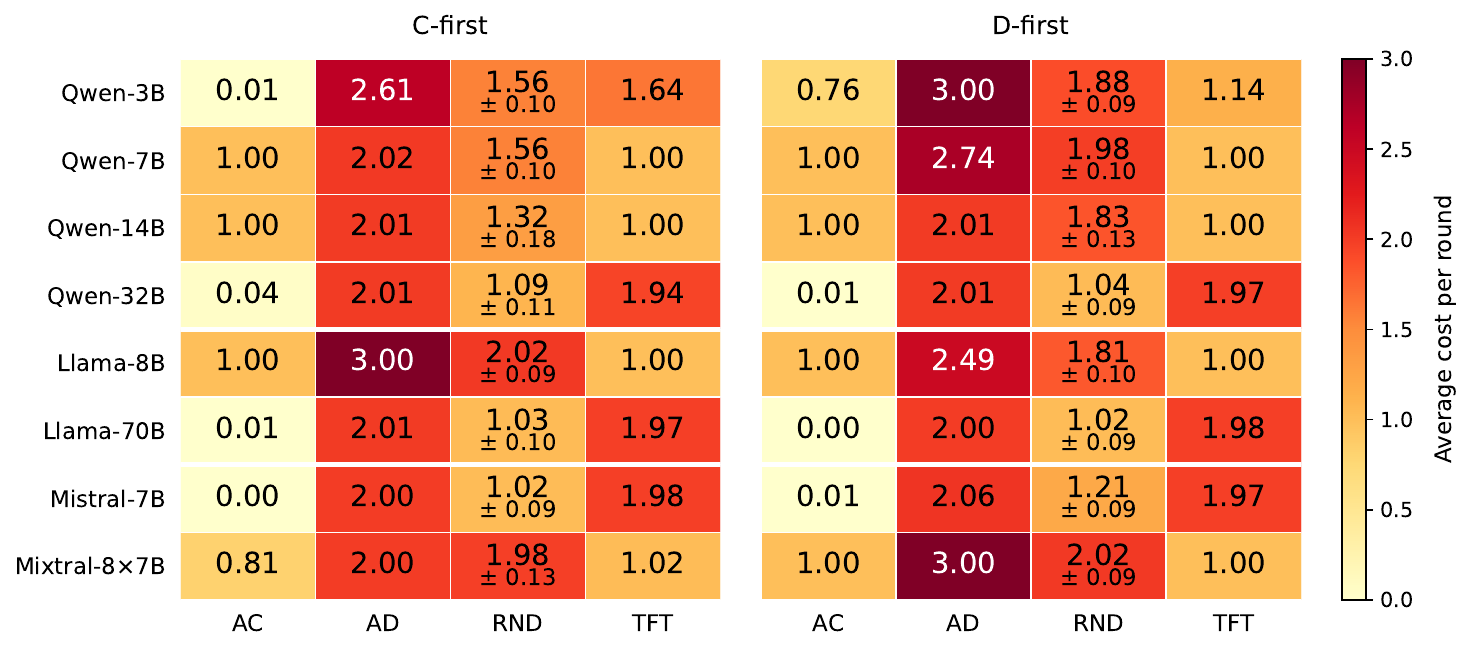}
    \caption{
    Individual average cost per round against fixed opponent
    policies under $C_{\mathrm{first}}$ and
    $D_{\mathrm{first}}$ prompting.
    Cells report means across seeds;
    non-zero sample standard deviations appear below.
    }
    \label{fig:dyadic-cost}
\end{figure}

\begin{figure}[tbp]
    \centering

    \begin{subfigure}{\linewidth}
        \centering
        \includegraphics[width=\linewidth]
        {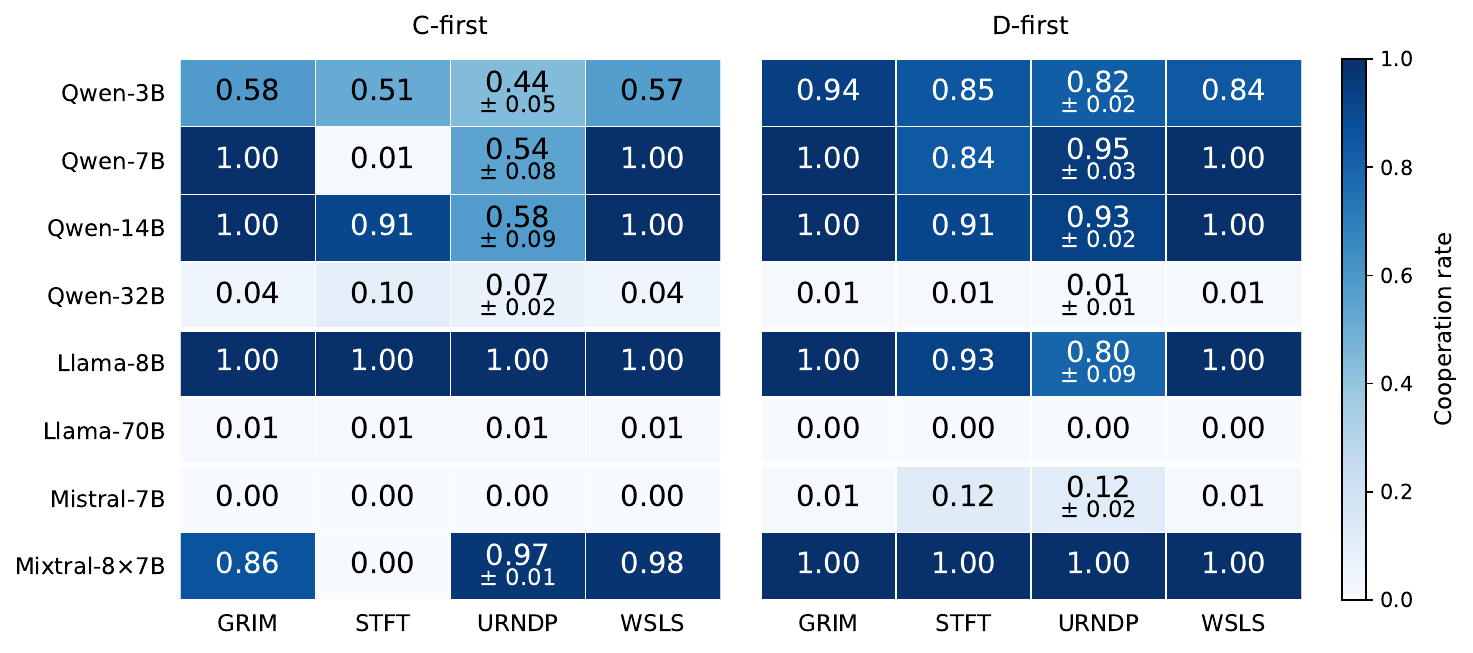}
        \caption{Cooperation rate.}
        \label{fig:dyadic-cooperation-app}
    \end{subfigure}

    \begin{subfigure}{\linewidth}
        \centering
        \includegraphics[width=\linewidth]
        {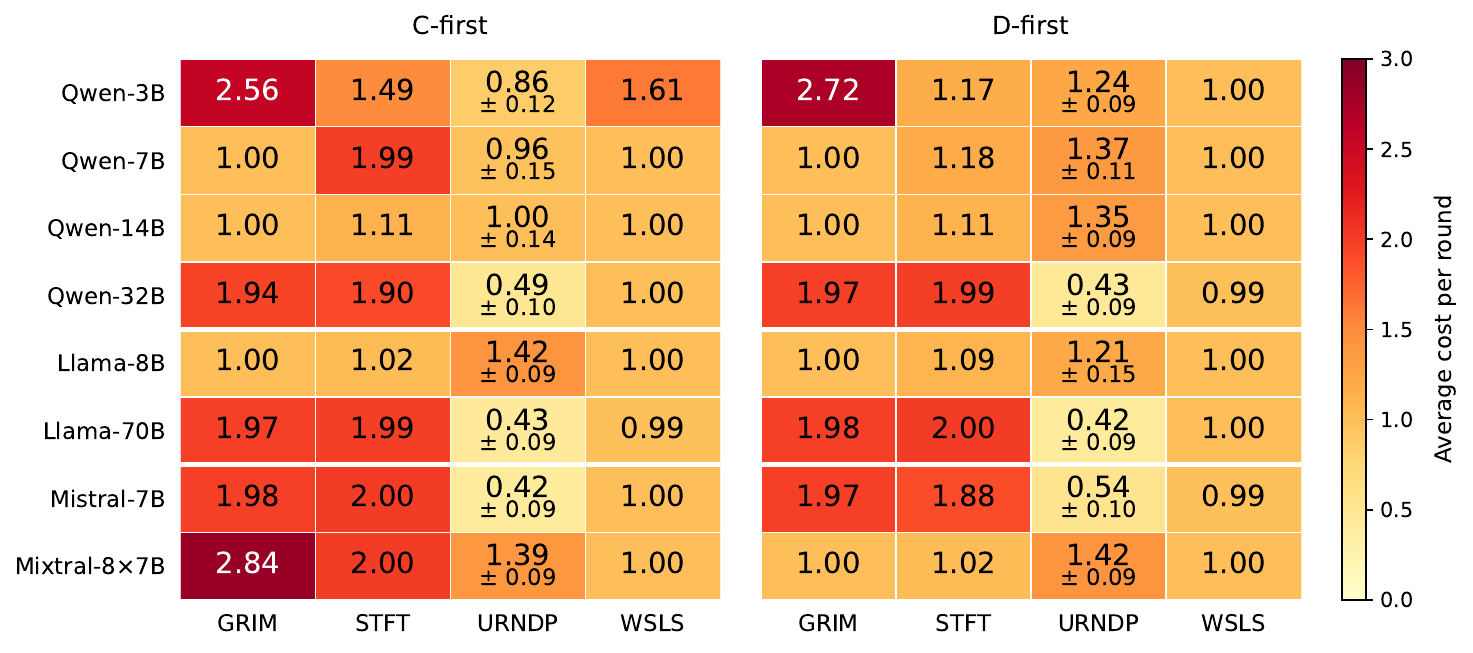}
        \caption{Individual cost per round.}
        \label{fig:dyadic-cost-app}
    \end{subfigure}

    \caption{
    Dyadic cooperation and individual cost against the
    remaining fixed policies under both label orders.
    Cells show seed means; nonzero sample SDs appear below.
    }
    \label{fig:dyadic-baselines-app}
\end{figure}

Figure~\ref{fig:dyadic-baselines-app} complements Figures~\ref{fig:dyadic-cooperation} and ~\ref{fig:dyadic-cost} with results against GRIM, STFT,
URNDP, and WSLS. Together, these figures distinguish cooperation
against a particular opponent from the individual cost resulting
from that interaction. Higher cooperation need not imply lower
cost, because the outcome also depends on the opponent's
responses. For example, under \texttt{C\_first}, Mixtral's
cooperation rate against GRIM is 0.86, yet its average cost is
2.84; under \texttt{D\_first}, full cooperation accompanies a
cost of 1.00. The paired panels therefore capture both behavioral
changes and their strategic consequences. Nonzero standard
deviations across seeds occur only against stochastic opponents,
consistent with variation in their sampled action sequences.

\begin{table}[t]
\centering
\scriptsize
\setlength{\tabcolsep}{1.7pt}
\renewcommand{\arraystretch}{1.12}

\resizebox{\columnwidth}{!}{%
\begin{tabular}{lcccccccc}
\toprule
\textbf{Scale}
& \textbf{AC}
& \textbf{AD}
& \textbf{GRIM}
& \textbf{RND}
& \textbf{STFT}
& \textbf{TFT}
& \textbf{URNDP}
& \textbf{WSLS} \\
\midrule

\multicolumn{9}{l}{\textbf{Qwen-2.5}} \\
\addlinespace[1pt]

3B
& $\underline{\mathbf{0.75}}$
& 0.39
& $\underline{0.36}$
& $0.32_{\pm .04}$
& 0.34
& $\underline{0.50}$
& $0.38_{\pm .05}$
& $\underline{0.27}$ \\

7B
& 0.00
& 0.72
& 0.00
& $0.43_{\pm .08}$
& \textbf{0.83}
& 0.00
& $\underline{0.41}_{\pm .06}$
& 0.00 \\

14B
& 0.00
& 0.00
& 0.00
& $\underline{\mathbf{0.51}}_{\pm .18}$
& 0.00
& 0.00
& $0.35_{\pm .08}$
& 0.00 \\

32B
& 0.03
& 0.00
& 0.03
& $0.05_{\pm .03}$
& \textbf{0.09}
& 0.03
& $0.06_{\pm .02}$
& 0.03 \\

\addlinespace[2pt]
\multicolumn{9}{l}{\textbf{Llama-3.1}} \\
\addlinespace[1pt]

8B
& 0.00
& \textbf{0.51}
& 0.00
& $0.22_{\pm .07}$
& 0.07
& 0.00
& $0.20_{\pm .09}$
& 0.00 \\

70B
& 0.01
& 0.01
& 0.01
& $0.01_{\pm .01}$
& 0.01
& 0.01
& 0.01
& 0.01 \\

\addlinespace[2pt]
\multicolumn{9}{l}{\textbf{Mistral}} \\
\addlinespace[1pt]

7B
& 0.01
& 0.06
& 0.01
& $\mathbf{0.19}_{\pm .03}$
& 0.12
& 0.01
& $0.12_{\pm .02}$
& 0.01 \\

8$\times$7B
& 0.19
& $\underline{\mathbf{1.00}}$
& 0.14
& $0.05_{\pm .04}$
& $\underline{\mathbf{1.00}}$
& 0.02
& $0.03_{\pm .01}$
& 0.02 \\

\bottomrule
\end{tabular}%
}

\caption{
Label-order sensitivity by model scale and opponent.
Bold marks row maxima (excluding fully tied rows);
underlining marks column maxima.
}
\label{tab:dyadic-si-by-opponent}
\end{table}

\section{Homogeneous Results}
\label{app:homogeneous-results}
Tables~\ref{tab:homogeneous_cooperation} and~\ref{tab:homogeneous_social_cost} provide the homogeneous uncertainty-off baselines
for each model, topology, network size, and label order.
Table~\ref{tab:homogeneous_cooperation} describes cooperation, while Table~\ref{tab:homogeneous_social_cost} reports its
collective cost consequences. Keeping configurations separate
reveals effects concealed by averages across network structures.
For example, Qwen-32B remains nearly fully cooperative in lines
and cycles under both orders, but its cooperation falls to 0.02
in larger stars under \texttt{D\_first}. Llama-70B also exhibits
different order effects in larger lines and cycles than in stars.

Social cost is normalized by the number of nodes and jointly evaluable rounds, but not by the number of edges. Consequently, cost differences across topologies reflect both strategic behavior and network connectivity. Comparisons within matched topology--size configurations are therefore more informative for isolating behavioral effects. Individual costs also depend on node degree, particularly in star networks. Finally, partial-round estimates characterize only their evaluable rounds and should not be interpreted as full-run outcomes.

\begin{table*}[t]
\centering
\small
\setlength{\tabcolsep}{4pt}
\renewcommand{\arraystretch}{1.00}
\resizebox{0.95\textwidth}{!}{%
\begin{tabular}{ll*{10}{c}c}
\toprule
\multirow{2}{*}{\textbf{Model}}
& \multirow{2}{*}{\textbf{Order}}
& \multicolumn{4}{c}{\textbf{Line}}
& \multicolumn{3}{c}{\textbf{Cycle}}
& \multicolumn{3}{c}{\textbf{Star}}
& \multirow{2}{*}{\textbf{Mean SI $\pm$ SD}} \\
\cmidrule(lr){3-6}
\cmidrule(lr){7-9}
\cmidrule(lr){10-12}
& & $N=3$ & $N=4$ & $N=10$ & $N=20$
& $N=4$ & $N=10$ & $N=20$
& $N=4$ & $N=10$ & $N=20$ & \\
\midrule

\multirow{2}{*}{Qwen-3B}
& C-first
& 0.00 & 0.00 & 0.00 & 0.00
& 0.00 & 0.00 & 0.00
& 0.00 & 0.00 & 0.00
& \multirow{2}{*}{$0.000 \pm 0.000$} \\
& D-first
& 0.00 & 0.00 & 0.00 & 0.00
& 0.00 & 0.00 & 0.00
& 0.00 & 0.00 & 0.00 & \\

\addlinespace[1pt]
\multirow{2}{*}{Qwen-7B}
& C-first
& 1.00 & 1.00 & 0.31 & 0.39
& 0.70 & 0.31 & 0.39
& 0.98 & 0.32 & 0.25
& \multirow{2}{*}{$0.435 \pm 0.320$} \\
& D-first
& 1.00 & 1.00 & 1.00 & 1.00
& 1.00 & 1.00 & 1.00
& 1.00 & 1.00 & 1.00 & \\

\addlinespace[1pt]
\multirow{2}{*}{Qwen-14B}
& C-first
& 1.00 & 1.00 & 1.00 & 1.00
& 1.00 & 1.00 & 1.00
& 1.00 & 1.00 & 1.00
& \multirow{2}{*}{$0.001 \pm 0.003$} \\
& D-first
& 1.00 & 1.00 & 1.00 & 1.00
& 1.00 & 1.00 & 1.00
& 1.00 & 0.99 & 1.00 & \\

\addlinespace[1pt]
\multirow{2}{*}{Qwen-32B}
& C-first
& 1.00 & 1.00 & 1.00 & 1.00
& 1.00 & 1.00 & 1.00
& 1.00 & 1.00 & 1.00
& \multirow{2}{*}{$0.197 \pm 0.413$} \\
& D-first
& 0.99 & 1.00 & 1.00 & 1.00
& 1.00 & 1.00 & 1.00
& 1.00 & 0.02 & 0.02 & \\

\midrule
\multirow{2}{*}{Llama-8B}
& C-first
& 0.99 & 0.99 & 0.96 & 0.94
& 0.99 & 0.93 & 0.92
& 1.00 & 1.00 & 1.00
& \multirow{2}{*}{$0.651 \pm 0.246$} \\
& D-first
& 0.15 & 0.13 & 0.50 & 0.53
& 0.56 & 0.50 & 0.55
& 0.14 & 0.09 & 0.06 & \\

\addlinespace[1pt]
\multirow{2}{*}{Llama-70B}
& C-first
& 0.78 & 0.94 & 0.20 & 0.11
& 0.96 & 0.08 & 0.05
& 0.51 & 0.87 & 0.91
& \multirow{2}{*}{$0.466 \pm 0.372$} \\
& D-first
& 0.84 & 0.93 & 0.96 & 0.97
& 0.98 & 0.98 & 0.98
& 0.87 & 0.52 & 0.50 & \\

\midrule
\multirow{2}{*}{Mistral-7B}
& C-first
& 1.00 & 1.00 & 1.00 & 1.00
& 1.00 & 1.00 & 1.00
& 1.00 & 1.00 & 1.00
& \multirow{2}{*}{$0.270 \pm 0.297$} \\
& D-first
& 0.67 & 1.00 & 0.85 & 0.88
& 1.00 & 0.75 & 0.93
& 0.01 & 0.51 & 0.70 & \\

\addlinespace[1pt]
\multirow{2}{*}{Mixtral-8$\times$7B}
& C-first
& 1.00 & 1.00 & 1.00 & 1.00
& 1.00 & 1.00 & 1.00
& 1.00 & 1.00 & 1.00
& \multirow{2}{*}{$0.000 \pm 0.000$} \\
& D-first
& 1.00 & 1.00 & 1.00 & \textemdash
& 1.00 & 1.00 & \textemdash
& 1.00 & 1.00 & 1.00 & \\

\bottomrule
\end{tabular}%
}
\caption{
Cooperation rates in homogeneous networks without the uncertainty cue.
Rates are averaged equally over nodes within each network,
then across three seeds; seed standard deviations are zero
and omitted.
Mean SI $\pm$ SD reports the mean and sample standard deviation
of absolute C-first--D-first cooperation differences across
topology--size configurations, weighted equally.
Dashes indicate unavailable rates because at least one node
has no valid action; configurations missing either ordering
are excluded from SI.
}
\label{tab:homogeneous_cooperation}
\end{table*}

\begin{table*}[t]
\centering
\small
\setlength{\tabcolsep}{4pt}
\renewcommand{\arraystretch}{1.00}
\resizebox{0.95\textwidth}{!}{%
\begin{tabular}{llcccccccccc}
\toprule
\textbf{Model} & \textbf{Order}
& \multicolumn{4}{c}{\textbf{Line}}
& \multicolumn{3}{c}{\textbf{Cycle}}
& \multicolumn{3}{c}{\textbf{Star}} \\
\cmidrule(lr){3-6}
\cmidrule(lr){7-9}
\cmidrule(lr){10-12}
& & $N=3$ & $N=4$ & $N=10$ & $N=20$
& $N=4$ & $N=10$ & $N=20$
& $N=4$ & $N=10$ & $N=20$ \\
\midrule

Qwen-3B & C-first
& 2.67 & 3.00 & 3.60 & 3.80
& 4.00 & 4.00 & 4.00 & 3.00 & 3.60 & 3.80 \\
& D-first
& 2.67 & 3.00 & 3.60 & 3.80
& 4.00 & 4.00 & 4.00 & 3.00 & 3.60 & 3.80 \\

\addlinespace[1pt]
Qwen-7B & C-first
& 1.33 & 1.51 & 3.09 & 3.07
& 2.60 & 3.39 & 3.22 & 1.52 & 2.49 & 2.67 \\
& D-first
& 1.33 & 1.50 & 1.80 & 1.90
& 2.00 & 2.00 & 2.00 & 1.50 & 1.80 & 1.90 \\

\addlinespace[1pt]
Qwen-14B & C-first
& 1.33 & 1.50 & 1.80 & 1.90
& 2.00 & 2.00 & 2.00 & 1.50 & 1.80 & 1.90 \\
& D-first
& 1.33 & 1.50 & 1.80 & 1.90
& 2.00 & 2.00 & 2.00 & 1.50 & 1.81 & 1.90 \\

\addlinespace[1pt]
Qwen-32B & C-first
& 1.33 & 1.50 & 1.80 & 1.90
& 2.00 & 2.00 & 2.00 & 1.50 & 1.80 & 1.90 \\
& D-first
& 1.35 & 1.51 & 1.81 & 1.90
& 2.01 & 2.00 & 2.00 & 1.50 & 3.56 & 3.75 \\

\midrule
Llama-8B & C-first
& 1.35 & 1.51 & 1.87 & 2.02
& 2.03 & 2.14 & 2.16 & 1.50 & 1.81 & 1.91 \\
& D-first
& 2.45 & 2.78 & 2.68 & 2.79
& 2.89 & 2.99 & 2.91 & 2.79 & 3.50 & 3.73 \\

\addlinespace[1pt]
Llama-70B & C-first
& 1.64 & 1.58 & 3.27 & 3.61
& 2.08 & 3.85 & 3.89 & 2.31 & 1.99 & 2.02 \\
& D-first
& 1.52 & 1.60 & 1.87 & 1.96
& 2.05 & 2.04 & 2.05 & 1.67 & 2.90 & 3.28 \\

\midrule
Mistral-7B & C-first
& 1.33 & 1.50 & 1.80 & 1.90
& 2.00 & 2.00 & 2.00 & 1.50 & 1.80 & 1.90 \\
& D-first
& 1.66 & 1.50 & $1.80^{\dagger}$ & $1.90^{\dagger}$
& $2.00^{\dagger}$ & $2.00^{\dagger}$ & $2.00^{\dagger}$
& 2.99 & $2.05^{\dagger}$ & 2.20 \\

\addlinespace[1pt]
Mixtral-8$\times$7B & C-first
& 1.33 & 1.50 & 1.80 & 1.90
& 2.00 & 2.00 & 2.00 & 1.50 & 1.80 & 1.90 \\
& D-first
& 1.33 & 1.51 & 1.81 & \textemdash
& 2.01 & 2.01 & \textemdash & 1.50 & 1.80 & 1.90 \\

\bottomrule
\end{tabular}%
}
\caption{
Mean social cost per node and jointly evaluable round,
$\overline{J}_{\mathrm{social}}$, in homogeneous networks
without the uncertainty cue. Entries are averaged across
three seeds; zero standard deviations are omitted.
$\dagger$ indicates estimates based on only one jointly
evaluable round per seed (two for Star $N=10$);
all other available estimates use 100 rounds.
Dashes indicate unavailable estimates.
}
\label{tab:homogeneous_social_cost}
\end{table*}

\section{Heterogeneous Results}
\label{app:het-results-unc-off}

\begin{figure}[tbp]
    \centering
    \includegraphics[width=\columnwidth]{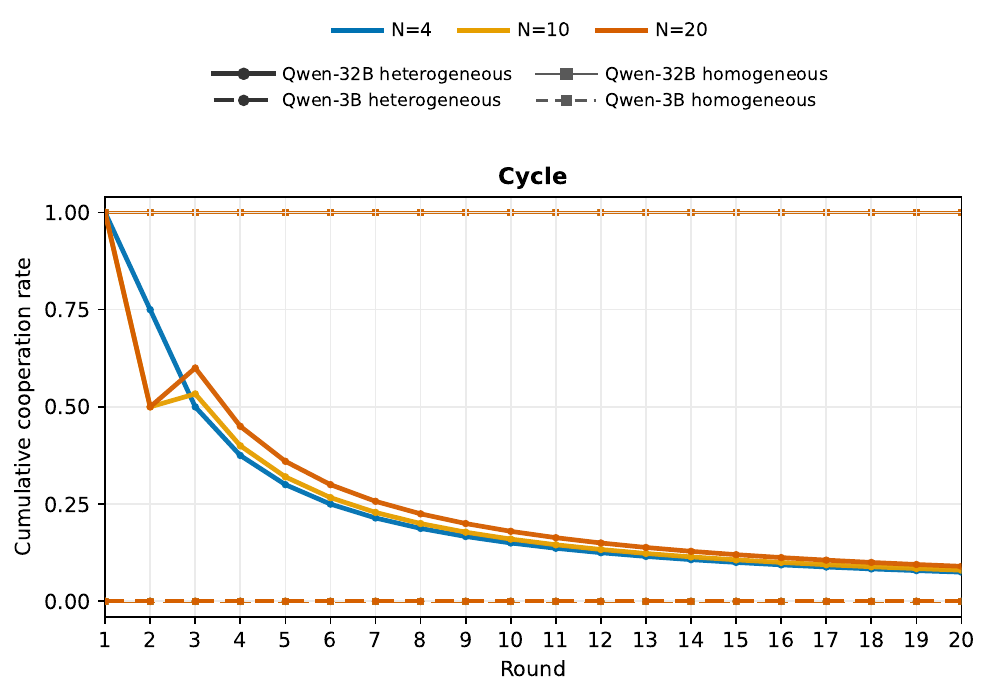}
        \caption{
    Cumulative cooperation in homogeneous and heterogeneous
    Qwen-3B/32B cycles under $\mathrm{C}_{\mathrm{first}}$
    without the uncertainty cue.
    }
    \label{fig:qwen_cycle_trajectories}
\end{figure}

\begin{figure*}[tp]
    \centering
    \begin{subfigure}[t]{0.48\textwidth}
        \centering
        \includegraphics[width=\linewidth]
        {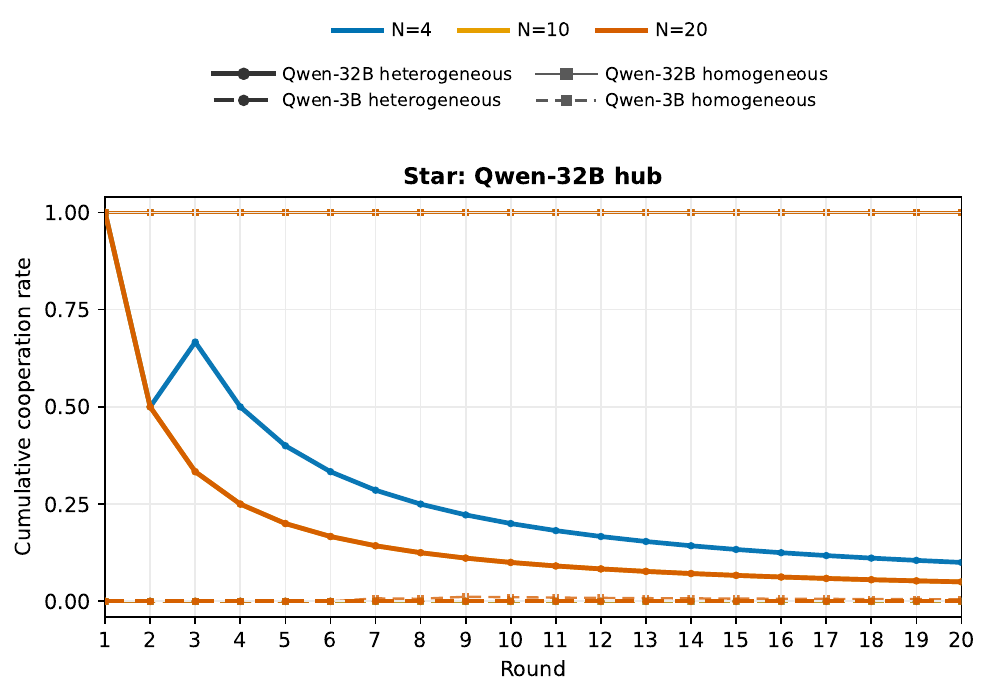}
        \caption{32B hub, 3B leaves.}
        \label{fig:qwen_star_large_hub_trajectories}
    \end{subfigure}
    \hfill
    \begin{subfigure}[t]{0.48\textwidth}
        \centering
        \includegraphics[width=\linewidth]
        {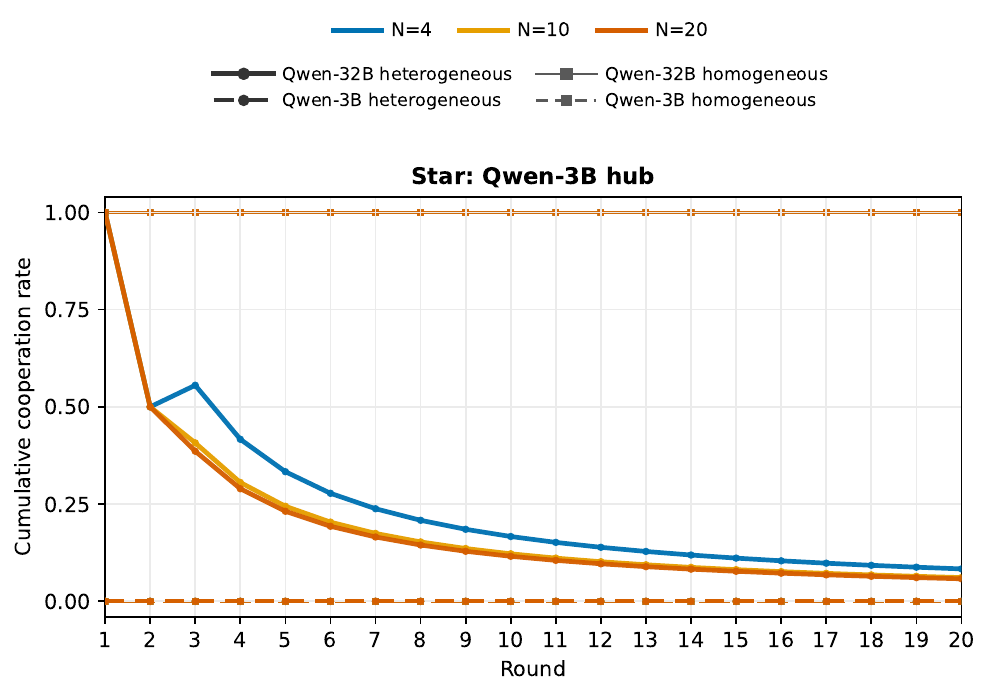}
        \caption{3B hub, 32B leaves.}
        \label{fig:qwen_star_small_hub_trajectories}
    \end{subfigure}
    \caption{
    Cumulative cooperation in homogeneous and heterogeneous
    Qwen-3B/32B stars under $\mathrm{C}_{\mathrm{first}}$
    without the cue. References match hub/leaf roles.
    }
    \label{fig:qwen_star_trajectories}
\end{figure*}

Figures~\ref{fig:qwen_cycle_trajectories} and~\ref{fig:qwen_star_trajectories} extend the trajectory comparison in Fig.~\ref{fig:qwen_line_trajectories} to
cycles and both star assignments for the Qwen-3B/32B pairing.
The homogeneous references show how each model behaves when
interacting with agents of its own model, with hub and leaf
roles matched in the star comparisons. Qwen-3B remains
predominantly defecting, while Qwen-32B's cooperation departs
from its homogeneous reference during heterogeneous interaction.

These figures display cumulative cooperation over the first
20 of 100 rounds. Each point summarizes actions from the
beginning of the run through the indicated round. A declining
curve therefore indicates lower cooperation in later decisions,
but does not necessarily imply a gradual, round-by-round
behavioral adjustment. The trajectories illustrate how
differences between homogeneous and heterogeneous networks
emerge during early interactions, while full-run cooperation
rates capture their overall magnitude
(Fig.~\ref{fig:heterogeneous_shifts}). The plots alone do not
establish whether these differences persist throughout all
100 rounds.

\begin{figure*}[tp]
    \centering
    \includegraphics[width=\textwidth]{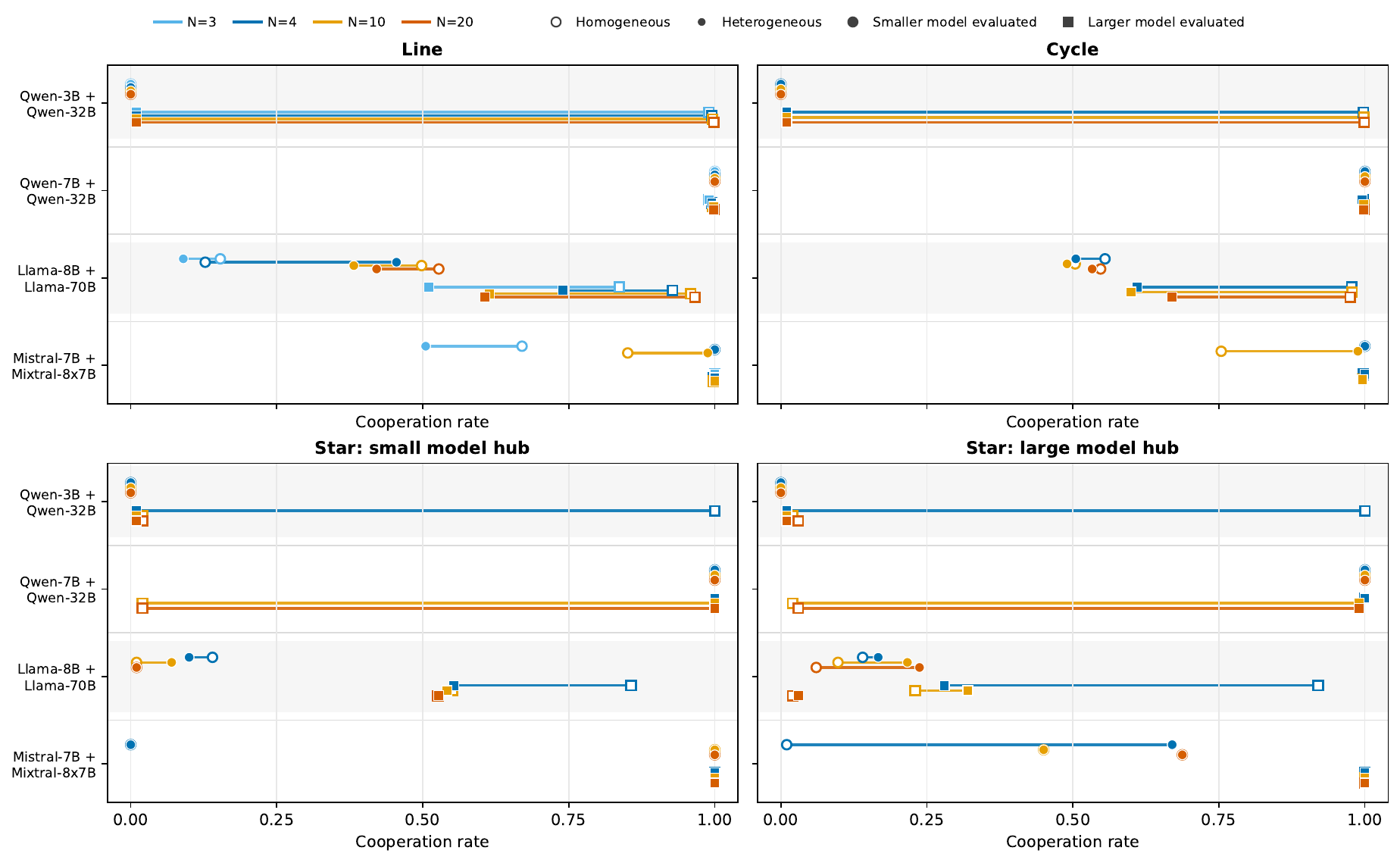}
    \caption{
    Cooperation shifts from homogeneous to heterogeneous graphs
    without the uncertainty cue ($D_{first}$). Lines connect matched settings;
    star baselines match hub/leaf roles. Values are averaged equally across seeds.
    }
    \label{fig:heterogeneous_shifts_dfirst}
\end{figure*}

Fig.~\ref{fig:heterogeneous_shifts_dfirst} complements
the $\mathrm{C}_{\mathrm{first}}$ comparison in
Fig.~\ref{fig:heterogeneous_shifts} by reporting
homogeneous-to-heterogeneous cooperation shifts under
$\mathrm{D}_{\mathrm{first}}$, without the uncertainty cue.
References are matched by model, topology, network size,
and seed, as well as hub/leaf role in stars.
Together, the two figures show which departures from
homogeneous behavior persist or change under label-order
reversal.

\section{Global Cue Exposure}
\label{app:global-uncertainty}
\subsection{Homogeneous Networks}
\label{app:global-uncertainty-homogeneous}
\begin{figure}[tbp]
    \centering
    \includegraphics[width=0.85\linewidth,
        trim={0 0 0 0},clip]{
        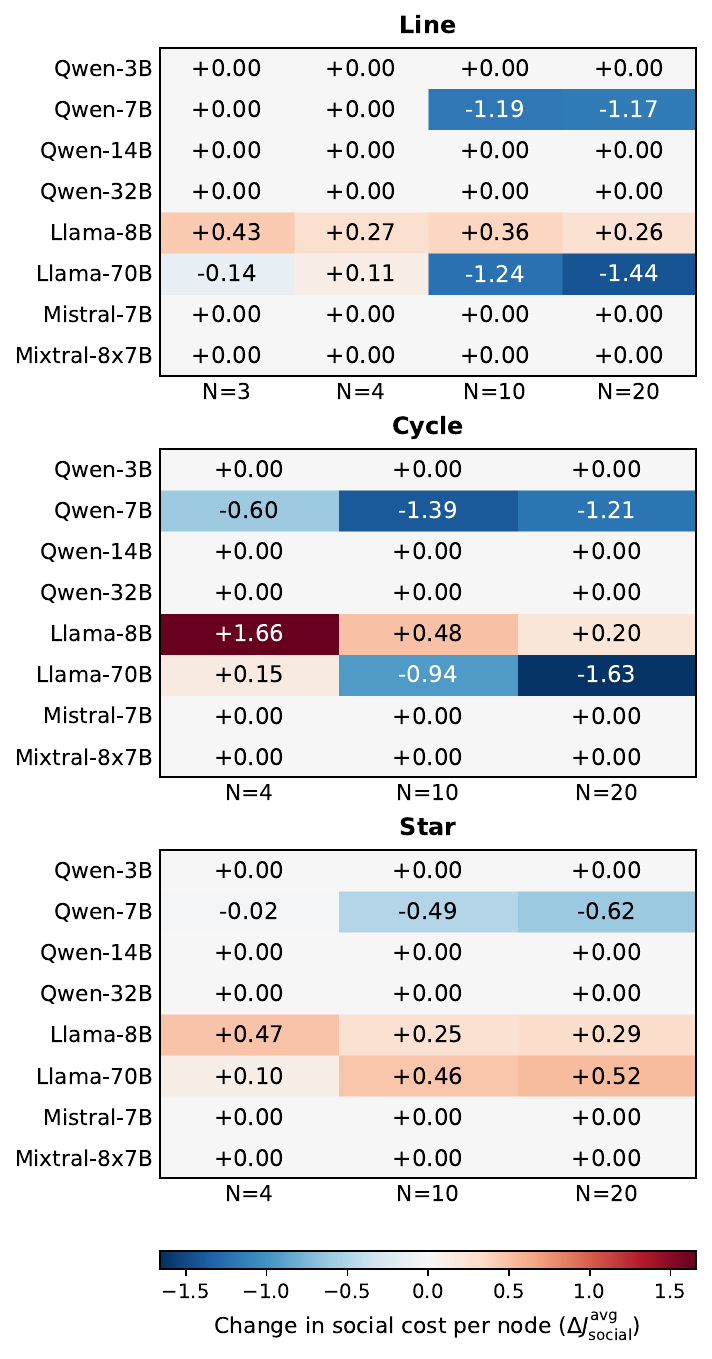
    }
    \caption{
Effect of global uncertainty cue on average social cost in
homogeneous networks under $\mathrm{C}_{\mathrm{first}}$, measured as
$\overline{J}_{\mathrm{social,on}}
- \overline{J}_{\mathrm{social,off}}$.
Blue indicates lower social cost (improved welfare);
red indicates higher social cost.
}
    \label{fig:homogeneous_social_uncertainty_effects}
\end{figure}

Fig.~\ref{fig:homogeneous_social_uncertainty_effects} reports the change in social cost when every agent receives the uncertainty cue, relative to matched configurations without the cue, while Table~\ref{tab:homogeneous-si-uncertainty-on} reports the SI for homogeneous networks under the global uncertainty cue exposure.

\begin{table}[t]
\centering
\small
\begin{tabular}{lc}
\toprule
\textbf{Model} & \textbf{Mean SI $\pm$ SD} \\
\midrule
Qwen-3B            & $0.0002 \pm 0.0003$ \\
Qwen-7B            & $0.0450 \pm 0.0762$ \\
Qwen-14B           & $0.9896 \pm 0.0046$ \\
Qwen-32B           & $0.0000 \pm 0.0000$ \\
Llama-8B           & $0.2414 \pm 0.2163$ \\
Llama-70B          & $0.3568 \pm 0.1412$ \\
Mistral-7B         & $0.0641 \pm 0.1088$ \\
Mixtral-8$\times$7B & $0.0000 \pm 0.0000$ \\
\bottomrule
\end{tabular}
\caption{
Label-order sensitivity in homogeneous networks under global exposure to the uncertainty cue. SI is the absolute cooperation-rate
difference between C-first and D-first, computed per
matched seed. Entries report the equally weighted mean
and sample SD across ten topology--size configuration
means, each averaged over three matched seeds.
}
\label{tab:homogeneous-si-uncertainty-on}
\end{table}

\subsection{Heterogeneous Networks}
\label{app:het-results-unc-on}

Table~\ref{tab:heterogeneous-global-uncertainty-summary}
shows that cooperation and welfare effects depend on
label order. Under $\mathrm{C}_{\mathrm{first}}$,
Qwen-7B/32B cooperation increases in both model groups,
while social cost falls by 0.268. However, Qwen-7B's
individual cost increases by 0.12, whereas Qwen-32B's
decreases by 2.75. In Llama-8B/70B networks, both models
become less cooperative under $\mathrm{C}_{\mathrm{first}}$,
with social cost increasing by 0.096. Under
$\mathrm{D}_{\mathrm{first}}$, their cooperation shifts
have opposite signs and social cost decreases by 0.030.

These comparisons show that collective improvements can
coexist with unequal individual consequences, and that
label order can alter the direction of uncertainty effects.
Averaging across configurations can further conceal
differences associated with network size, topology, and
hub assignment. However, pooled averages may also mask
configuration-specific reversals, as illustrated
in Table~\ref{tab:heterogeneous-global-uncertainty-configurations}.

Table~\ref{tab:heterogeneous-global-uncertainty-configurations}
complements these averages with selected configuration-specific comparisons. Panel~(a) illustrates cost redistribution between
model groups, reversals across network sizes, and effects that disappear under a different label order or hub assignment. Panel~(b) compares heterogeneous cooperation with matched uncertainty-off homogeneous references, showing a reduced absolute gap in selected Qwen-32B and Llama-8B configurations.
\begin{table*}[t]
\centering
\small
\setlength{\tabcolsep}{3pt}
\renewcommand{\arraystretch}{1.12}
\resizebox{\textwidth}{!}{%
\begin{tabular}{@{}lllrrrrrr@{}}
\toprule
Pair & Model & Order
& \shortstack{Cooperation\\off $\to$ on}
& $\Delta$ coop.
& \shortstack{Individual cost\\off $\to$ on}
& $\Delta$ cost
& \shortstack{Social cost/node\\off $\to$ on}
& $\Delta$ social \\
\midrule

\multirow{4}{*}{\shortstack[l]{Llama-8B\\+ Llama-70B}}
& \multirow{2}{*}{Llama-8B}
& C-first
& $0.698 \to 0.588$ & $-0.110$
& $5.34 \to 5.41$ & $+0.07$
& $2.442 \to 2.538$ & $+0.096$ \\
& & D-first
& $0.283 \to 0.491$ & $+0.208$
& $3.86 \to 6.17$ & $+2.31$
& $2.811 \to 2.781$ & $-0.030$ \\
& \multirow{2}{*}{Llama-70B}
& C-first
& $0.560 \to 0.524$ & $-0.036$
& $5.13 \to 5.18$ & $+0.05$
& $2.442 \to 2.538$ & $+0.096$ \\
& & D-first
& $0.508 \to 0.373$ & $-0.134$
& $6.37 \to 4.62$ & $-1.74$
& $2.811 \to 2.781$ & $-0.030$ \\

\addlinespace
\multirow{4}{*}{\shortstack[l]{Mistral-7B\\+ Mixtral-8$\times$7B}}
& \multirow{2}{*}{Mistral-7B}
& C-first
& $1.000 \to 1.000$ & $0.000$
& $3.55 \to 3.55$ & $0.00$
& $1.764 \to 1.764$ & $0.000$ \\
& & D-first
& $0.790 \to 0.973$ & $+0.182$
& $3.44 \to 3.83^{\dagger}$ & $+0.40$
& $1.942 \to 1.744^{\dagger}$ & $-0.198$ \\
& \multirow{2}{*}{Mixtral-8$\times$7B}
& C-first
& $1.000 \to 1.000$ & $0.000$
& $3.63 \to 3.63$ & $0.00$
& $1.764 \to 1.764$ & $0.000$ \\
& & D-first
& $1.000 \to 1.000$ & $0.000$
& $6.48 \to 4.48^{\dagger}$ & $-2.00$
& $1.942 \to 1.744^{\dagger}$ & $-0.198$ \\

\addlinespace
\multirow{4}{*}{\shortstack[l]{Qwen-3B\\+ Qwen-32B}}
& \multirow{2}{*}{Qwen-3B}
& C-first
& $0.000 \to 0.000$ & $0.000$
& $7.01 \to 6.47$ & $-0.54$
& $3.515 \to 3.395$ & $-0.120$ \\
& & D-first
& $0.000 \to 0.000$ & $0.000$
& $7.04 \to 7.04$ & $0.00$
& $3.519 \to 3.519$ & $0.000$ \\
& \multirow{2}{*}{Qwen-32B}
& C-first
& $0.015 \to 0.147$ & $+0.133$
& $7.31 \to 7.58$ & $+0.28$
& $3.515 \to 3.395$ & $-0.120$ \\
& & D-first
& $0.010 \to 0.010$ & $0.000$
& $7.30 \to 7.30$ & $0.00$
& $3.519 \to 3.519$ & $0.000$ \\

\addlinespace
\multirow{4}{*}{\shortstack[l]{Qwen-7B\\+ Qwen-32B}}
& \multirow{2}{*}{Qwen-7B}
& C-first
& $0.755 \to 0.971$ & $+0.216$
& $3.41 \to 3.52$ & $+0.12$
& $2.059 \to 1.791$ & $-0.268$ \\
& & D-first
& $1.000 \to 1.000$ & $0.000$
& $3.56 \to 3.55$ & $-0.01$
& $1.767 \to 1.764$ & $-0.003$ \\
& \multirow{2}{*}{Qwen-32B}
& C-first
& $0.933 \to 1.000$ & $+0.067$
& $7.15 \to 4.40$ & $-2.75$
& $2.059 \to 1.791$ & $-0.268$ \\
& & D-first
& $0.996 \to 1.000$ & $+0.004$
& $3.60 \to 3.63$ & $+0.03$
& $1.767 \to 1.764$ & $-0.003$ \\

\bottomrule
\end{tabular}%
}
\caption{
Global uncertainty-cue effects in heterogeneous networks,
reported separately by label order.
Off/on denote cue absence/presence; $\Delta$ denotes on minus off.
Model-level metrics average nodes equally within each model group,
then matched seeds, followed by equal weighting of configurations.
Social cost weights model groups by node count within each run
and is reported per node and jointly evaluable round;
network values are repeated for both models.
Each order uses 13 configurations per pairing, except D-first
Mixtral cooperation and Mistral/Mixtral costs, which use 11.
$^{\dagger}$ marks costs containing two partial-coverage
configurations in the off condition.
All remaining cost estimates use 100 jointly evaluable rounds
per run. Differences use unrounded values.
}
\label{tab:heterogeneous-global-uncertainty-summary}
\end{table*}

\begin{table}[t]
\centering
\small
\setlength{\tabcolsep}{3pt}
\begin{tabular}{@{}llcc@{}}
\toprule
\textbf{Pair} & \textbf{Model}
& \textbf{SI: cue off} & \textbf{SI: cue on} \\
\midrule

\multirow{4}{*}{\shortstack[l]{Llama-8B\\+ Llama-70B}}
& Llama-8B  & \sisd{0.435}{0.353} & \sisd{0.296}{0.266} \\[3pt]
& Llama-70B & \sisd{0.387}{0.239} & \sisd{0.286}{0.227} \\

\addlinespace
\multirow{4}{*}{\shortstack[l]{Mistral-7B\\+ Mixtral-8$\times$7B}}
& Mistral-7B & \sisd{0.227}{0.321} & \sisd{0.030}{0.056} \\[3pt]
& Mixtral-8$\times$7B & \sisd{0.000}{0.001} & \sisd{0.000}{0.000} \\

\addlinespace
\multirow{4}{*}{\shortstack[l]{Qwen-3B\\+ Qwen-32B}}
& Qwen-3B  & \sisd{0.000}{0.000} & \sisd{0.000}{0.000} \\[3pt]
& Qwen-32B & \sisd{0.005}{0.004} & \sisd{0.137}{0.148} \\

\addlinespace
\multirow{4}{*}{\shortstack[l]{Qwen-7B\\+ Qwen-32B}}
& Qwen-7B  & \sisd{0.245}{0.290} & \sisd{0.029}{0.073} \\[3pt]
& Qwen-32B & \sisd{0.066}{0.107} & \sisd{0.000}{0.000} \\

\bottomrule
\end{tabular}
\caption{
Label-order sensitivity in heterogeneous networks without
and with the global uncertainty cue.
SI is the absolute cooperation-rate difference between
C-first and D-first, computed per matched configuration
and seed after equally averaging nodes within each model group.
Entries report mean $\pm$ sample SD across equally weighted
configuration means, each averaged over matched seeds.
Only configurations with eligible estimates in both cue
conditions are included.
}
\label{tab:heterogeneous-si-global-uncertainty}
\end{table}

\begin{table*}[t!]
\centering
\small
\setlength{\tabcolsep}{4pt}
\textbf{(a) Selected configuration-specific effects}\par\smallskip
\begin{tabular}{lrrrrrr}
\toprule
Model & $C_{\mathrm{off}}\to C_{\mathrm{on}}$ & $\Delta C$
& $I_{\mathrm{off}}\to I_{\mathrm{on}}$ & $\Delta I$
& $S_{\mathrm{off}}\to S_{\mathrm{on}}$ & $\Delta S$ \\
\midrule
\multicolumn{7}{l}{\textit{Qwen-3B/Qwen-32B; Cycle, $N=4$, alternating; \texttt{C\_first}}} \\
Qwen-3B & $0.000\to 0.000$ & $0.000$ & $3.940\to 2.160$ & $-1.780$ & \multirow{2}{*}{$3.985\to 3.540$} & \multirow{2}{*}{$-0.445$} \\
Qwen-32B & $0.015\to 0.460$ & $+0.445$ & $4.030\to 4.920$ & $+0.890$ &  &  \\
\addlinespace
\multicolumn{7}{l}{\textit{Qwen-3B/Qwen-32B; Cycle, $N=4$, alternating; \texttt{D\_first}}} \\
Qwen-3B & $0.000\to 0.000$ & $0.000$ & $3.960\to 3.960$ & $0.000$ & \multirow{2}{*}{$3.990\to 3.990$} & \multirow{2}{*}{$0.000$} \\
Qwen-32B & $0.010\to 0.010$ & $0.000$ & $4.020\to 4.020$ & $0.000$ &  &  \\
\addlinespace
\multicolumn{7}{l}{\textit{Llama-8B/Llama-70B; Cycle, $N=4$, alternating; \texttt{C\_first}}} \\
Llama-8B & $0.910\to 0.175$ & $-0.735$ & $4.200\to 2.910$ & $-1.290$ & \multirow{2}{*}{$2.685\to 3.465$} & \multirow{2}{*}{$+0.780$} \\
Llama-70B & $0.405\to 0.360$ & $-0.045$ & $1.170\to 4.020$ & $+2.850$ &  &  \\
\addlinespace
\multicolumn{7}{l}{\textit{Llama-8B/Llama-70B; Cycle, $N=10$, alternating; \texttt{C\_first}}} \\
Llama-8B & $0.400\to 0.486$ & $+0.086$ & $4.400\to 3.204$ & $-1.196$ & \multirow{2}{*}{$3.500\to 3.072$} & \multirow{2}{*}{$-0.428$} \\
Llama-70B & $0.100\to 0.442$ & $+0.342$ & $2.600\to 2.940$ & $+0.340$ &  &  \\
\addlinespace
\multicolumn{7}{l}{\textit{Qwen-7B/Qwen-32B; Star, $N=20$, 32B hub; \texttt{C\_first}}} \\
Qwen-7B & $0.259\to 0.842$ & $+0.583$ & $0.299\to 0.842$ & $+0.543$ & \multirow{2}{*}{$2.623\to 2.050$} & \multirow{2}{*}{$-0.573$} \\
Qwen-32B & $0.980\to 1.000$ & $+0.020$ & $46.760\to 25.000$ & $-21.760$ &  &  \\
\addlinespace
\multicolumn{7}{l}{\textit{Qwen-7B/Qwen-32B; Star, $N=20$, 7B hub; \texttt{C\_first}}} \\
Qwen-7B & $1.000\to 1.000$ & $0.000$ & $19.000\to 19.000$ & $0.000$ & \multirow{2}{*}{$1.900\to 1.900$} & \multirow{2}{*}{$0.000$} \\
Qwen-32B & $1.000\to 1.000$ & $0.000$ & $1.000\to 1.000$ & $0.000$ &  &  \\
\addlinespace
\bottomrule
\end{tabular}

\medskip
\textbf{(b) Movement toward matched homogeneous references}\par\smallskip
\begin{tabular}{llrrr}
\toprule
Model & Heterogeneous configuration & $C_H$
& $C_{\mathrm{off}}\to C_{\mathrm{on}}$
& $|C-C_H|$: off $\to$ on \\
\midrule
Qwen-32B & Qwen-3B/Qwen-32B, cycle, $N=4$ & $1.000$ & $0.015\to 0.460$ & $0.985\to 0.540$ \\
Llama-8B & Llama-8B/Llama-70B, cycle, $N=10$ & $0.932$ & $0.400\to 0.486$ & $0.532\to 0.446$ \\
\bottomrule
\end{tabular}
\caption{Selected effects of global exposure to the uncertainty cue in heterogeneous networks.
(a) $C$ denotes model-group cooperation, $I$ mean individual cost,
and $S$ social cost per node; changes are on minus off.
Social cost is shared by the two model rows and uses node-count
weighting within each run.
(b) Both comparisons use \texttt{C\_first}; $C_H$ is the
uncertainty-off homogeneous cooperation rate matched by model,
topology, size, and label order. A smaller absolute gap indicates
movement toward this reference.
All entries average three seeds, with no averaging across sizes,
compositions, or label orders. All displayed cost estimates use
100 jointly evaluable rounds per run. Seed SDs are zero and omitted;
differences are calculated before rounding.}
\label{tab:heterogeneous-global-uncertainty-configurations}
\end{table*}

\section{Local Cue Exposure}
\label{app:local-uncertaity}

\begin{figure*}[tp]
    \centering
    \includegraphics[width=\textwidth]
    {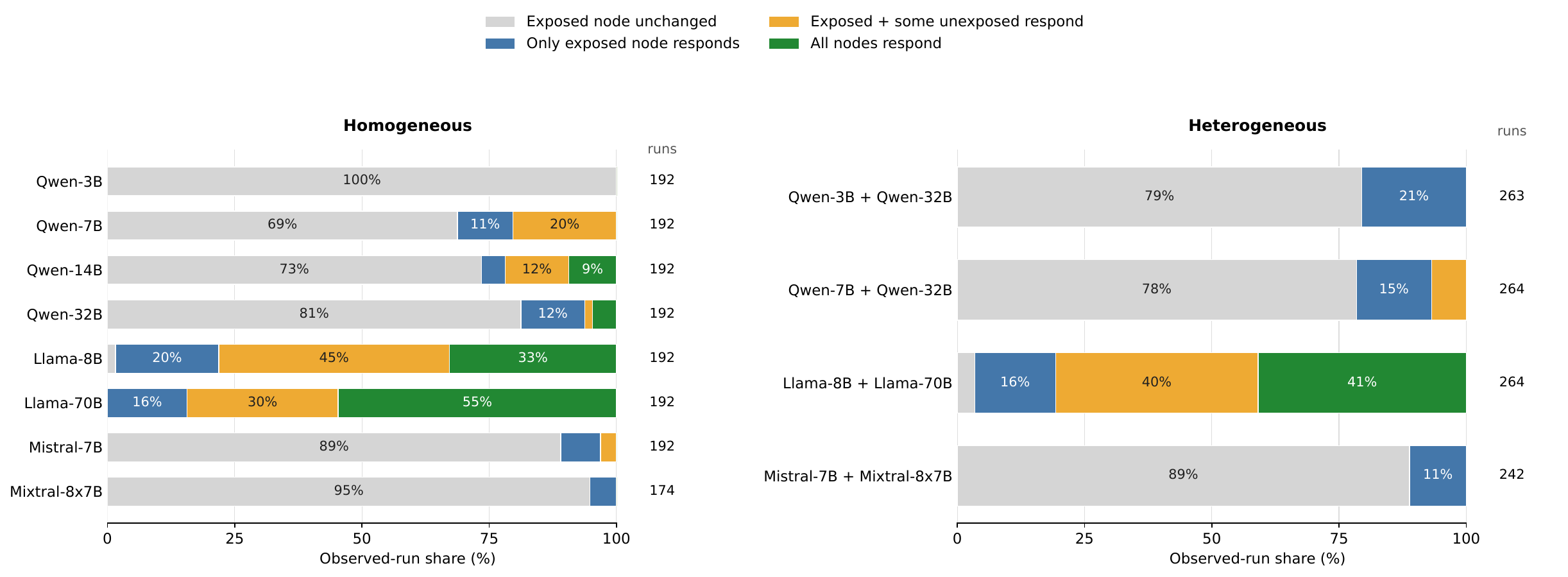}
    \caption{
Responses to localized uncertainty relative to matched
cue-off runs. Categories distinguish no exposed-node action
response and exposed-node responses with zero, partial,
or full temporal reach among unexposed nodes.
Reach is based on action differences linked through time
along network edges, not proof of causal transmission.
Bars report observed-run shares pooled across configurations,
label orders, and seeds; counts give included runs.
}

\label{fig:localized_response_extent}
\end{figure*}

Figure~\ref{fig:localized_response_extent} reveals substantial differences in the extent of local intervention effects across model families. Multi-agent responses are particularly prevalent in Llama networks: heterogeneous
Llama-8B/70B configurations exhibit exposed-node action
responses with partial temporal reach in approximately
40\% of observed runs and full reach in another 41\%. Other model pairings more frequently exhibit
no exposed-node action response, although this does not rule
out changes elsewhere. These proportions are descriptive, pooling network structures, sizes, exposure locations, label orders, and seeds. 

\paragraph{Interpretation of response categories.}
Fig.~\ref{fig:localized_response_extent} classifies responses
according to round-by-round action differences and their
temporal reach. Importantly, the ``exposed node unchanged''
category does not imply that unexposed agents remain unaffected.
Consequently, the displayed shares cannot directly establish
the overall prevalence of behavioral spillovers, as these may
also occur without an exposed-node action response.
Counts include three seed repetitions per configuration,
which typically yield identical trajectories under greedy
decoding. These percentages are therefore descriptive rather
than estimates based on independent observations.


Table~\ref{tab:localized_uncertainty_roles}  examines whether the response remains concentrated at the exposed node or extends to the surrounding network, and how its benefits and costs are distributed. The Qwen-3B/32B example shows that unexposed agents can benefit through lower costs without changing their cooperation. Conversely, the Qwen-7B hub example shows that a small mean response at the exposed node can accompany substantial changes elsewhere. Comparing hub and leaf exposures in the same Llama network further illustrates the importance of the exposed node's position and model identity. Collective gains can also coexist with higher individual costs for some agents. These averages summarize tendencies across sizes; the magnitude and direction of effects may differ between individual configurations.

\begin{table*}[t]
\centering
\small
\setlength{\tabcolsep}{2.8pt}
\renewcommand{\arraystretch}{1.10}
\resizebox{\textwidth}{!}{%
\begin{tabular}{@{}llllrrrrrr@{}}
\toprule
Network & Topology/composition & Order & Exposed node
& $\Delta C_E$ & $\Delta C_U$ & $A_U$
& $\Delta I_E$ & $\Delta I_U$ & $\Delta S$ \\
\midrule

Qwen-3B/32B & Cycle, alternating & C-first & 32B node
& 0.400 & 0.000 & 0.000 & 0.800 & -0.265 & -0.107 \\
& & D-first & 32B node
& 0.000 & 0.000 & 0.000 & 0.000 & 0.000 & 0.000 \\

\addlinespace
Llama-8B/70B & Star, 70B hub & C-first & 70B hub
& 0.203 & 0.411 & 0.481 & -5.783 & 0.004 & -0.458 \\
& & D-first & 70B hub
& -0.127 & -0.144 & 0.182 & 2.140 & 0.109 & 0.228 \\
& & C-first & 8B leaf
& -0.186 & 0.009 & 0.010 & -0.183 & 0.034 & 0.010 \\
& & D-first & 8B leaf
& 0.210 & 0.109 & 0.212 & -0.168 & -0.262 & -0.226 \\

\addlinespace
Llama-8B/70B & Star, 8B hub & C-first & 8B hub
& -0.157 & -0.108 & 0.143 & -0.617 & 0.205 & 0.211 \\
& & D-first & 8B hub
& 0.197 & 0.067 & 0.073 & 0.217 & -0.326 & -0.202 \\
& & C-first & 70B leaf
& -0.134 & -0.022 & 0.108 & -0.079 & 0.090 & 0.050 \\
& & D-first & 70B leaf
& -0.278 & 0.050 & 0.054 & -0.500 & -0.001 & -0.078 \\

\addlinespace
Qwen-7B & Star, homogeneous & C-first & Hub
& 0.010 & 0.269 & 0.274 & -7.183 & 0.249 & -0.257 \\
& & D-first & Hub
& 0.000 & 0.000 & 0.000 & 0.000 & 0.000 & 0.000 \\

\addlinespace
Llama-8B & Cycle, homogeneous & C-first & Cycle node
& -0.239 & -0.023 & 0.035 & -0.360 & 0.234 & 0.121 \\
& & D-first & Cycle node
& -0.012 & -0.057 & 0.142 & 0.276 & 0.067 & 0.098 \\

\addlinespace
Qwen-32B & Star, homogeneous & C-first & Hub
& 0.000 & 0.000 & 0.000 & 0.000 & 0.000 & 0.000 \\
& & D-first & Hub
& 0.650 & 0.653 & 0.653 & -9.210 & -0.647 & -1.206 \\

\bottomrule
\end{tabular}%
}
\caption{
Selected effects of localized cue exposure relative to
matched cue-off controls, averaged equally over $N=4,10,20$.
$\Delta C_E$ and $\Delta C_U$ denote mean cooperation changes
for exposed and unexposed nodes;
$A_U$ averages absolute node-level cooperation changes
before aggregation.
$\Delta I_E$ and $\Delta I_U$ denote mean individual-cost
changes, and $\Delta S$ denotes social-cost change per node.
Matched seeds are averaged within exposure locations,
locations within roles, and sizes equally.
Label orders are reported separately.
}
\label{tab:localized_uncertainty_roles}
\end{table*}

\paragraph{Computational resources.}
Experiments were conducted on the Leonardo Booster
partition at CINECA. Each compute node provides four
NVIDIA A100 GPUs (Ampere architecture), each with
64\,GiB of HBM2e memory and NVLink~3.0 connectivity,
alongside a 32-core Intel Xeon Platinum 8358 CPU
at 2.6\,GHz and 512\,GB of system memory.

\end{document}